\documentclass[english,aps,pra,notitlepage,twocolumn,secnumarabic]{revtex4-1}

\usepackage{graphicx}
\usepackage{amsmath}
\usepackage{dsfont}
\usepackage{xspace}
\usepackage[usenames,dvipsnames]{color}
\usepackage{stmaryrd}
\usepackage[normalem]{ulem}
\usepackage{comment}

\DeclareMathOperator{\tr}{\mbox{tr}}
\DeclareMathOperator{\re}{\mbox{Re}}

\definecolor{dgreen}{rgb}{0.0, 0.5, 0.0}

\newcommand{\eq}[1]{Eq.\thinspace{}(\ref{#1})}

\newcommand{\Eq}[1]{Eq.\thinspace{}(\ref{#1})}

\newcommand{\fig}[1]{Fig.\thinspace{}\ref{#1}}

\newcommand{\fc}[1]{({#1})}

\def\ket#1{\mathinner{|{#1}\rangle}}

\def\beq{\begin{equation}}
\def\eeq{\end{equation}}
\def\bea{\begin{eqnarray}}
\def\eea{\end{eqnarray}}

\newcommand{\veck}{\mathbf k}

\newcommand{\vect}[1]{{\mathbf #1}}

\definecolor{dgreen}{rgb}{0.0, 0.5, 0.0}

\begin{document}

\title{Ultrafast many-body interferometry of impurities coupled to a Fermi sea}

\author{M. Cetina$^{1, 2}$, M. Jag$^{1, 2}$, R. S. Lous$^{1, 2}$, I. Fritsche$^{1, 2}$, J. T. M. Walraven$^{1, 3}$, R. Grimm$^{1, 2}$, \\
J. Levinsen$^{4}$, M. M. Parish$^{4}$, R. Schmidt$^{5, 6}$, M. Knap$^{7}$,  E. Demler$^{6}$}

\affiliation{$^{1}$Institut f\"ur Quantenoptik und Quanteninformation, \"Osterreichische Akademie der Wissenschaften, 6020 Innsbruck, Austria}

\affiliation{$^{2}$Institut f\"ur Experimentalphysik, Universit\"at Innsbruck, 6020 Innsbruck, Austria}

\affiliation{$^{3}$Van der Waals-Zeeman Institute, Institute of Physics, University of Amsterdam, 1098 XH Amsterdam, The Netherlands}

\affiliation{$^{4}$School of Physics and Astronomy, Monash University, Victoria 3800, Australia}

\affiliation{$^{5}$ITAMP, Harvard-Smithsonian Center for Astrophysics, Cambridge, MA 02138, USA} 

\affiliation{$^{6}$Department of Physics, Harvard University, Cambridge, MA 02138, USA}

\affiliation{$^{7}$Department of Physics, Walter Schottky Institute and Institute for Advanced Study, Technical University of Munich, 85748 Garching, Germany}


\date{\today}

\begin{abstract}
The fastest possible collective response of a quantum many-body system is related to its excitations at the 
highest possible energy. In condensed-matter systems, the corresponding timescale is typically set by the Fermi energy.
Taking advantage of fast and precise control of interactions between 
ultracold atoms, we report on the observation of ultrafast dynamics of impurities coupled to an atomic 
Fermi sea. Our interferometric measurements track the non-perturbative quantum evolution
of a fermionic many-body system, revealing in real time the formation dynamics of quasiparticles and 
the quantum interference between attractive and repulsive states throughout the 
full depth of the Fermi sea. Ultrafast time-domain methods to manipulate and investigate strongly 
interacting quantum gases open up new windows on the dynamics of quantum matter under extreme 
non-equilibrium conditions.
\end{abstract}

\maketitle

\hyphenation{Fesh-bach}

Non-equilibrium dynamics of fermionic systems is at the heart of many problems in science and technology, from the physics of neutron stars and heavy ion collisions to the operation of electronic devices. 
The wide range of energy scales, spanning the low energies of excitations near the Fermi surface up to high energies of excitations from deep within the Fermi sea, challenges our understanding of the quantum dynamics 
in such fundamental systems.
The Fermi energy $E_F$ sets the shortest response time for the collective response of a fermionic many-body system through the Fermi time $\tau_{F} = \hbar/E_{F}$, where $\hbar$ is the reduced Planck constant. In a metal, i.e. a Fermi sea of electrons, $E_{F}$ is in the range of a few electronvolts, which corresponds to $\tau_{F}$ on the order of 100 attoseconds. Dynamics in condensed matter systems on this timescale can be recorded by attosecond streaking techniques \cite{Krausz2009apx} and the initial applications were demonstrated by probing photoelectron emission from a surface \cite{Pazourek2015aco}. However, despite these spectacular advances, the direct observation of the coherent evolution of a fermionic many-body system on the Fermi timescale has remained beyond reach.

In atomic quantum gases, the fermions are much heavier and the densities far lower, which brings $\tau_{F}$ into the experimentally accessible range of typically a few microseconds. 
Furthermore, the powerful techniques of atom interferometry \cite{Cronin2009oai} now offer 
the exciting opportunity to probe and manipulate the real-time coherent evolution of a fermionic quantum many-body system. 
Such techniques 
have been successfully used, e.g. to measure bosonic Hanbury-Brown-Twiss correlations \cite{Simon2011qso}, demonstrate topological bands \cite{Atala2013dmo}, probe quantum and thermal fluctuations in low-dimensional condensates \cite{Gring2012rap, Hadzibabic2006bkt}, and to measure demagnetization dynamics of a fermionic gas  \cite{Koschorreck2013usd,Bardon2014tdd}. 
Impurities coupled to a quantum gas provide a novel and unique probe of the many-body state. Strikingly, they allow direct access to the system's wave function when the internal states of the impurities are manipulated using a Ramsey atom-interferometric technique \cite{Goold2011oca,Knap2012tdi}.

We employ dilute $^{40}$K atoms in a $^{6}$Li Fermi sea to measure the response of the sea to a suddenly introduced impurity. For near-resonant interactions, we observe coherent quantum many-body dynamics involving the entire $^{6}$Li Fermi sea. We also observe in real time the formation dynamics of the repulsive and attractive impurity quasiparticles. In the limit of low impurity concentration, our experiments confirm that an elementary Ramsey sequence is equivalent to linear-response frequency-domain spectroscopy. We demonstrate that our time-domain approaches allow us to prepare, control, and measure many-body interacting states.

\begin{figure}[t]
\begin{center}
\includegraphics[width=9cm]{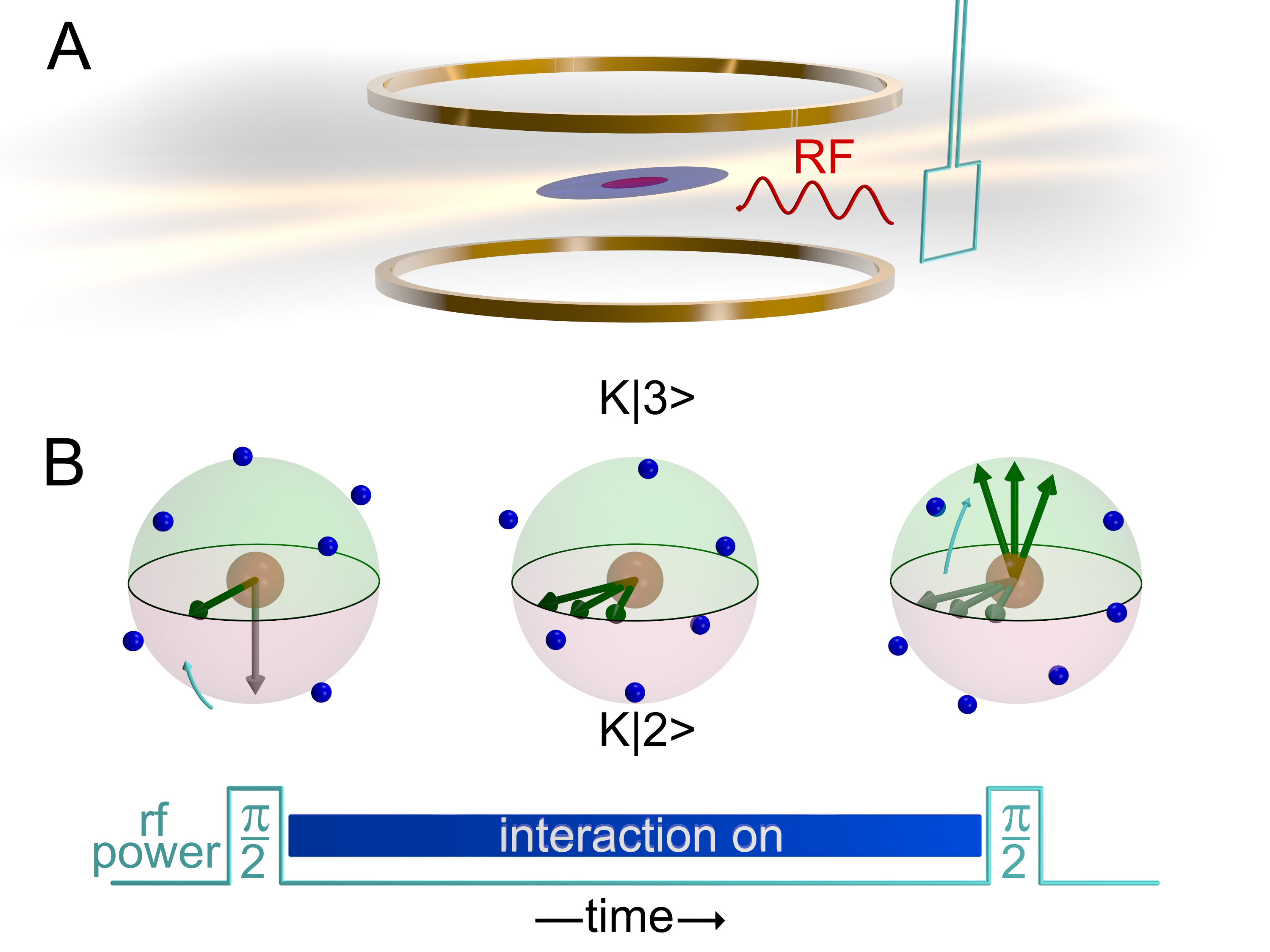}
\end{center}
\caption{Illustration of the experimental setup and procedure. 
(\textbf{A}) Li (blue) and K (red) atoms are held in a crossed-beam optical dipole trap.
The magnetic field coils (gold) and the rf coil (blue) are used to manipulate the atoms.
(\textbf{B}) An rf $\pi/2$ pulse is used to prepare the K atoms in a superposition of internal
Zeeman states as shown on a Bloch sphere. A second rf pulse is used to probe the final state.
}
\end{figure}

Our system consists of a small sample of typically $1.5\times 10^{4}$ $^{40}$K impurity atoms immersed in a Fermi sea of 3$\times$$10^{5}$ $^{6}$Li atoms \cite{Cetina2015doi,supmat}.  The mixture is held in an optical dipole trap (Fig.~1A) at a temperature of $T = 430\,$nK after forced evaporative cooling.  Because of the Li Fermi pressure and a more than two times stronger optical potential for K, the K impurities are concentrated in the central region of the large Li cloud. Here they experience a nearly homogeneous environment with an effective Fermi energy of $\epsilon_{F} = k_B \times 2.6 \mu$K \cite{supmat}, where $k_B$ is Boltzmann's constant. The corresponding Fermi time $\tau_F = 2.9~\mu$s sets the natural time scale for our experiments. The degeneracy of the Fermi sea is characterized by $k_{B}T/\epsilon_{F} \approx 0.17$. The concentration of K in the Li sea remains low, with $\bar{n}_{{\rm K}}/\bar{n}_{{\rm Li}} \approx 0.2$, where $\bar{n}_{\rm Li}$ ($\bar{n}_{\rm K}$) is the average of the Li (K) number density sampled by the K atoms \cite{supmat}. 

The interaction between the impurity atoms in the internal state K$|3\rangle$ (third-to-lowest Zeeman sublevel) and the Li atoms (always kept in the lowest Zeeman sublevel) is controlled using a rather narrow \cite{supmat} interspecies Feshbach resonance near a magnetic field of 154.7~G \cite{Naik2011fri,Cetina2015doi}. We quantify the interaction with the Fermi sea by the dimensionless parameter $X \equiv -1/\kappa_{F}a$, where $\kappa_{F} = \hbar^{-1}\sqrt{2m_{\rm Li}\epsilon_{F}}$ is the Li Fermi wavenumber with $m_{\rm Li}$ the Li mass, and $a$ is the $s$-wave interspecies scattering length.  While slow control of $X$ is realized in a standard way by variations of the magnetic field, fast control is achieved using an optical resonance shifting technique \cite{Cetina2015doi}. The latter permits sudden changes of $X$ by up to about $\pm 5$ within a time shorter than $\tau_F/15 \approx 200$~ns.

Our interferometric probing method is based on a two-pulse Ramsey
scheme (Fig.~1B), following the suggestions of Refs.~\cite{Goold2011oca,Knap2012tdi}.
The sequence starts with the impurity atoms prepared in the spin state
K$|2\rangle$ (second-to-lowest Zeeman sublevel), for which the 
background interaction with the Fermi sea can be neglected. 
A first, 10-$\mu$s-long, radio-frequency (rf) $\pi/2$-pulse drives the K atoms into a coherent
superposition between this non-interacting initial state and the state
K$|3\rangle$ under weakly interacting conditions
(interaction parameter $X_{1}$ with $\left|X_{1}\right|\approx 5$).
Using the optical resonance
shifting technique \cite{Cetina2015doi}, the system is then rapidly
quenched into the strongly interacting regime ($\left|X\right|<1$).
After an evolution time $t$, the system is quenched back into the
regime of weak interactions and a second 
$\pi/2$-pulse is applied. The population difference $N_{3}-N_{2}$
in the two impurity states is measured as a function of the phase 
of the rf pulse \cite{Cetina2015doi}. The contrast
$|S(t)|$ and the phase $\varphi$ of the resulting sinusoidal signal
is finally determined as a function of $t$.
In the limit of low impurity concentration, the complex 
function $S(t)=|S(t)|e^{-i\varphi(t)}$ 
can be interpreted as the overlap of the interacting 
and the non-interacting components of the system's wavefunction \cite{Goold2011oca}.
The squared amplitude $|S(t)|^{2}$ is then equivalent 
to the common definition of a Loschmidt echo \cite{Loschmidt1876udz,Jalalbert2001tst}.

\begin{figure*}[tbh]
\centering
\includegraphics[width=16cm]{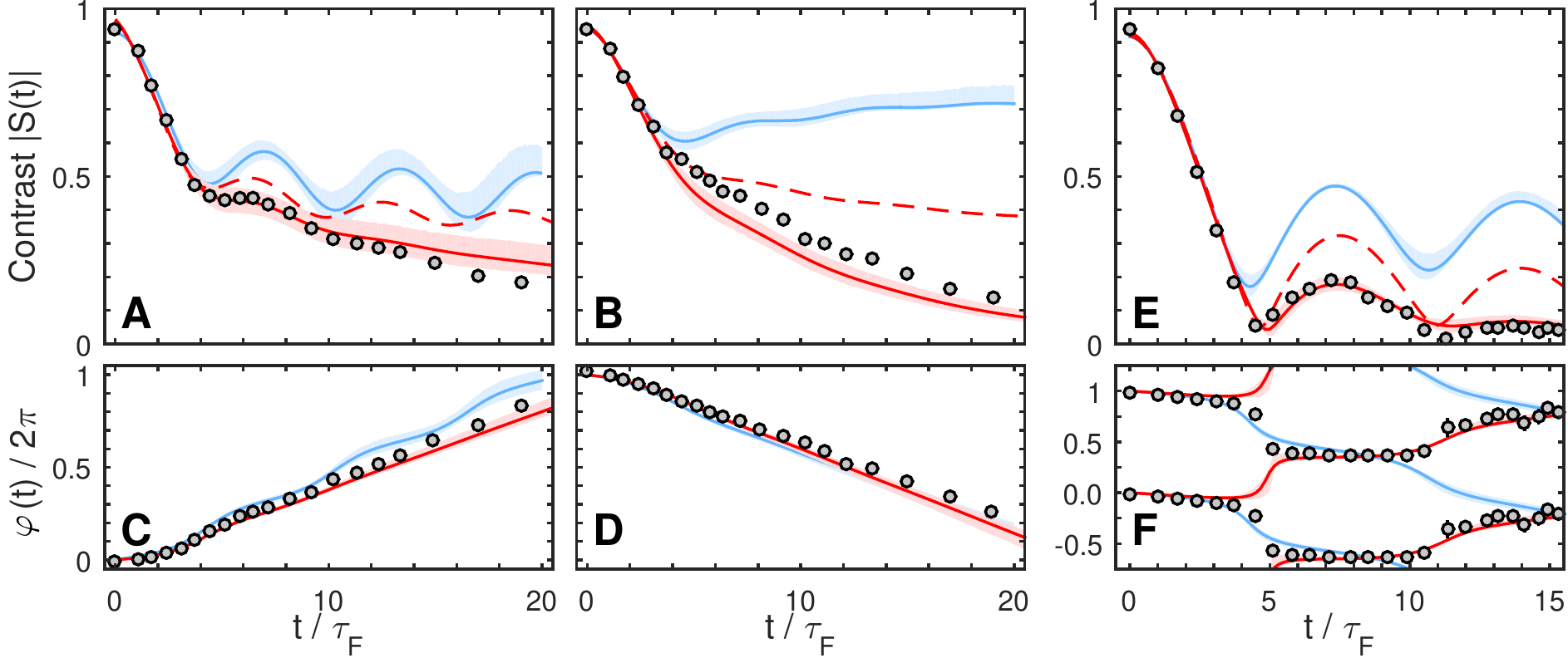}
\caption{Impurity dynamics in the Fermi sea. 
(\textbf{A} and \textbf{C}) Contrast $|S(t)|$ and phase $\varphi(t)$
of the interference signal depending on the interaction time $t$ in the
repulsive polaron regime for $X=-0.23(6)$, with the rf pulse applied at $X_1=-3.9$.
(\textbf{B} and \textbf{D}) Same quantities in the attractive polaron regime for $X=0.86(6)$ and $X_1=5.8$.
(\textbf{E} and \textbf{F}) Same quantities for resonant interactions ($X=0.08(5)$, $X_1=4.8$).
The solid blue lines show the results of the TBM calculations.
The solid (dashed) red lines show the results of the FDA calculations at the measured (at zero) temperature.
The shaded regions reflect the combined experimental uncertainties in $X$, $k_B T$ and $\epsilon_F$. 
The errors in the experimental data are typically smaller than the symbol size.}
\end{figure*}

We first consider the interaction conditions where polaronic quasiparticles are known to exist \cite{Massignan2014pdm}. 
Figures 2A-D show the evolution of the contrast and the phase measured in the repulsive and the attractive polaron regimes,
where $X=-0.23(6)$ and $X=+0.86(6)$, respectively. For short evolution times of up to about $4\tau_{F}$, we observe
both contrast signals to exhibit a similar initial parabolic transient, which is typical of a Loschmidt echo \cite{Jalalbert2001tst}. 
For longer times, this connects to an exponential decay of the contrast and a linear evolution of the phase. 
In Ref.~\cite{Cetina2015doi}, we showed that the long-time decay of the contrast in this regime can be interpreted in terms of quasiparticle scattering.
Here, the linear phase evolution corresponds to the energy shift of the quasiparticle state, for which we obtain $+0.29(1)\epsilon_{F}$
for the repulsive case in Fig.~2C and $-0.27(1)\epsilon_{F}$ for the attractive case in Fig.~2D.
Remarkably, while the long-time behavior reflects the quasiparticle properties, 
the observed initial transient reveals the ultrafast real-time dynamics of the quasiparticle formation.

On resonance, for the strongest possible interactions, the quasiparticle picture breaks down. 
Here our measurements, displayed in Fig.~2E and 2F for $X$ = 0.08(5), reveal the striking quantum dynamics
of a strongly interacting fermionic system forced into an extreme non-equilibrium state.
The contrast $|S(t)|$ shows pronounced oscillations reaching zero, while the phase $\varphi(t)$ exhibits plateaus. 
The revivals of the contrast $|S(t)|$ indicate partially reversible entanglement between the internal state of the impurity and the Fermi sea \cite{Goold2011oca}.
This process involves the whole Fermi sea and occurs on the fastest timescale available to the collective dynamics of a fermionic system.

To further interpret our measurements we employ two different theoretical approaches:
the truncated basis method (TBM) \cite{supmat} and the functional determinant approach (FDA) \cite{Knap2012tdi}.
The TBM models our full experimental procedure assuming zero temperature and considering only single particle-hole excitations. 
This approximation, known as the Chevy ansatz \cite{Chevy2006upd}, has been successfully used to predict
the properties of quasiparticles in cold gases \cite{Massignan2014pdm}. 
The predictions of the TBM are represented by the blue lines in Fig.~2. 
This method accurately describes the initial transient, as well as the period of the oscillations of $S(t)$ on resonance.
While the zero-temperature TBM calculation naturally overestimates the contrast in the thermally dominated regime ($t>6\tau_F$),
it accurately reproduces the observed linear phase evolution and thus the quasiparticle energy.
The FDA is an exact solution for a fixed impurity at arbitrary temperatures taking into account the non-perturbative creation of infinitely many particle-hole pairs. 
The FDA calculation is represented by the solid red lines in Fig.~2. 
We see remarkable agreement with our experimental results, which indicates that the effects of impurity motion remain small in our system.
This observation can be explained by the fact that our impurity is sufficiently heavy so that 
the effects of its recoil with energies of about $0.25\,\epsilon_{F}$ \cite{supmat} are masked by thermal fluctuations.
To identify the effect of temperature, we performed a corresponding FDA 
calculation for $T=0$ and show the results as the dashed lines in Fig. 2.
Here, we see a slower decay of $|S(t)|$, which follows a power law at long times \cite{supmat} under the idealizing assumption of infinitely heavy impurities.

\begin{figure*}[tbh]
\begin{center}
\includegraphics[width=16cm]{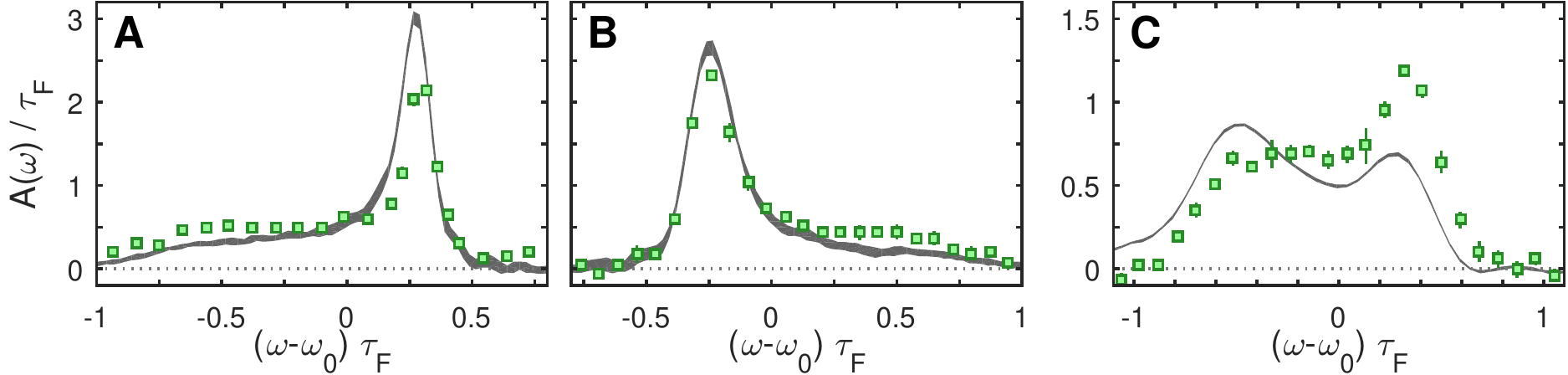}
\end{center}
\caption{Rf spectroscopy of  an impurity in the Fermi sea.
(\textbf{A} and \textbf{B}) show the rf spectra for the repulsive ($X=-0.23(6)$) and the attractive ($X=0.86(6)$) interactions, respectively.
(\textbf{C}) shows the rf spectrum for resonant interactions ($X=0.08(5)$).
The spectral data are normalized to unit integral.
The gray lines correspond to the numerical Fourier transform of the $S(t)$ data from Fig. 2.
The width of the gray curve reflects the combined experimental errors in the $S(t)$ data.
}
\end{figure*}

Time-domain and frequency-domain methods are closely related, as is well known in spectroscopy. 
In the limit of low impurity density, where the interactions between the impurities can be neglected,
$S(t)$ is predicted to be proportional to the inverse Fourier transform of the linear excitation spectrum 
$A\left(\omega\right)$ of the impurity \cite{Nozieres1969sit}. To benchmark our interferometric method,
we measure $A\left(\omega\right)$ using rf spectroscopy similar to our earlier work \cite{Kohstall2012mac},
but with great care to ensure linear response \cite{supmat}.
The measured excitation spectra are shown in Fig.~3. 
In the repulsive and attractive polaron regimes, we observe the characteristic structure of a peak on top of a broad pedestal \cite{Massignan2014pdm}.  
While the peak determines the long-time evolution of the quasiparticle, the pedestal is associated with the rapid dynamics related to the emergence of many-body correlations.
For resonant interactions, the rf response is broad and nearly symmetric about $\omega_0$, implying that the zero crossings of $S(t)$ are accompanied by jumps in its phase by $\pi$, as is seen in Fig.~2E and 2F.
Based on the observed spectral response, we interpret the oscillations of $S(t)$ in Figs.~2E and 2F 
as arising from simultaneous excitations of the two branches of our many-body system corresponding to the two humps in the rf spectrum.

A detailed comparison of our time- and frequency-domain measurements reveals
the powerful capability of our approach to prepare and control many-body states. 
This is revealed in Fig.~3, where we show the Fourier transform of the $S\left(t\right)$ data from Fig.~2 as the gray curves. 
We observe that time-domain measurements where the rf pulses are applied
in the presence of weakly repulsive interactions (Fig.~3A) emphasize the upper branch
of the many-body system while in the attractive case (Fig.~3B,C), the lower branch is emphasized.
We explain this observation by the action of the rf pulses 
to prepare weakly interacting polaron states \cite{supmat}.
Compared to the non-interacting initial state used in the frequency-domain spectroscopy,
these polarons have an increased wavefunction overlap with the corresponding strongly interacting repulsive and attractive branches, 
leading to the observed shift in the spectral weight.
Our measurements demonstrate that the control over the initial state 
of many particles can be used to precisely manipulate quantum dynamics in the strongly interacting regime. 
This unique capability of time-domain techniques opens up a wide range of applications, including the study of the dynamical behavior near the phase 
transition from a polaronic to a molecular system \cite{Massignan2014pdm} and the creation of specific excitations of a Fermi 
sea down to individual atoms \cite{Dubois2013mes}.

\begin{figure}[h]
\begin{center}
\includegraphics[width=7cm]{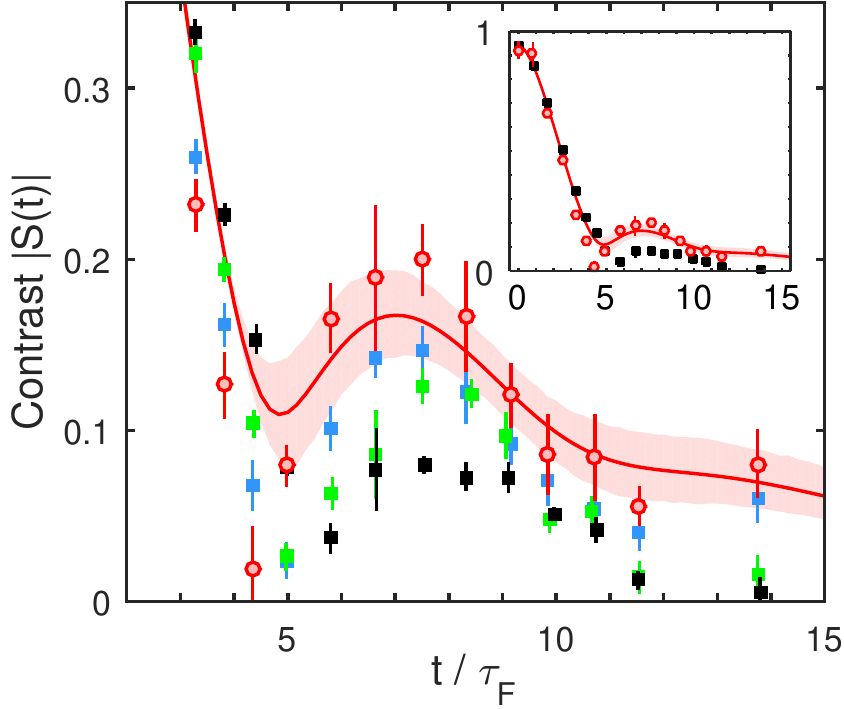}
\end{center}
\caption{Observation of induced impurity-impurity interactions. 
Resonant dynamics of the contrast is shown for $X=-0.01(5)$, $X_1=5.2$, 
$\epsilon_F = k_B\times2.1(1)$ $\mu$K, 
$k_B T/\epsilon_F = 0.24(2)$ and different impurity concentrations $\bar{n}_{{\rm K}}/\bar{n}_{{\rm Li}}$. 
The black, green, and blue squares correspond to $\bar{n}_{{\rm K}}/\bar{n}_{{\rm Li}}=0.53$, $0.33$, and $0.20$, respectively.
The red circles correspond to the linear extrapolation of the complex $S(t)$ data to the limit of a single impurity,
taking into account the errors in the data. The inset reproduces this extrapolation together with the highest-concentration data points.
The red line shows the result of the FDA calculation. The shaded region reflects the combined experimental uncertainties in $X$, $k_B T$ and $\epsilon_F$. 
}
\end{figure}

Our interpretation of the results in Figs.~2 and 3 relies on the assumption 
that our fermionic impurities are sufficiently dilute so that any interactions between them can be neglected.
We can extend our experiments into a complex many-body regime where the impurities interact 
both with the Fermi sea and with each other, by increasing the impurity concentration \cite{supmat}. 
Figure 4 shows the time-dependent contrast measured for $k_B T=0.24(2) \epsilon_F$ and 
$\bar{n}_{\rm{K}}/\bar{n}_{\rm{Li}}=$0.20, 0.33, and 0.53. 
An extrapolation of the $S(t)$ data to zero concentration (open red circles) lies close to the data
points for $\bar{n}_{\rm{K}}/\bar{n}_{\rm{Li}}=$0.20, which is the typical
concentration in our measurements, and agrees with the FDA calculation. 
This confirms that the physics that we 
access in the measurements with a small sample of fermionic impurities is close to that of a single impurity, 
which we posit to be a consequence of the fermionic nature of the impurities. 
When the impurity concentration is increased, we find that the contrast for $t>5\tau_F$ is decreased and 
the period of the revivals of $|S(t)|$ is prolonged. We interpret this as arising from effective interactions between
the impurities induced by the Fermi sea \cite{Mora2010npo,Yu2010con}.
Such interactions between fermionic impurities are predicted to lead to novel quantum phases \cite{Zwerger2012tbb}.

Our results demonstrate the power of many-body interferometry to study ultrafast processes in strongly interacting Fermi gases in real time,
including the formation dynamics of quasiparticles and the extreme non-equilibrium dynamics arising from quantum interference between different many-body branches.
Of particular interest is the prospect of observing Anderson's orthogonality catastrophe \cite{Knap2012tdi,supmat} by
further cooling the Li Fermi sea \cite{Hart2015ooa} while pinning the K atoms in a deep species-selective optical lattice\cite{LeBlanc2007sso}.

\begin{center}
\textbf{Acknowledgements}
\end{center}
We thank M. Baranov, F. Schreck, G. Bruun, N. Davidson and R. Folman for stimulating discussions. We acknowledge support by the Austrian Science Fund FWF within the SFB FoQuS (F40-P04). R.~S. was supported by the NSF through a grant for the Institute for Theoretical Atomic, Molecular, and Optical Physics at Harvard University and the Smithsonian Astrophysical Observatory. M.~K. acknowledges support from Technical University of Munich - Institute for Advanced Study, funded by the German Excellence Initiative and the European Union FP7 under grant agreement 291763. E.~D. acknowledges support from Harvard-MIT CUA, NSF Grant No. DMR-1308435, AFOSR Quantum Simulation MURI, the ARO-MURI on Atomtronics, and support from  Dr.~Max R\"ossler, the Walter Haefner Foundation, the ETH Foundation, and the Simons Foundation.

\onecolumngrid
\setlength{\abovecaptionskip}{6pt plus 3pt minus 2pt}
\setlength{\belowcaptionskip}{0pt}
\vspace{15mm}

\begin{center}
\textbf{\large Supplementary Materials}
\end{center}

\tableofcontents

\section{Theoretical description}
In this section, we summarize the approaches that we developed
to theoretically model the results of our interferometric Ramsey
experiments.  We first discuss the microscopic model that
we use to describe the narrow Feshbach resonance of
the Li-K mixture, and then we outline how we
calculate the time evolution of the system within two approaches: the
Truncated Basis Method (TBM) and the Functional Determinant Approach
(FDA).  In this section, we assume that a `perfect quench' is
performed, where the impurity is initially non-interacting with the
Fermi sea and there are no interactions during the radio-frequency
(rf) pulses.  A discussion of the role played by
interactions during the rf pulses is deferred to Section \ref{sec:S3}.

\subsection{Narrow Feshbach resonance model for Li-K mixtures}

In our experiment, the K impurities are concentrated in the central
region of the Li Fermi gas where they experience a nearly uniform Li
environment (see Section S5.A).  Hence we consider in our model K
impurities that are immersed in a Li Fermi gas of uniform density.
The Li-K mixture is prepared at magnetic fields near a closed-channel
dominated Feshbach resonance between the Li$|1\rangle$ and
K$|3\rangle$ states that occurs near $155$ G. The narrow character of
this resonance is a consequence of the limited strength of the
coupling of atoms in the open channel to a closed-channel molecular
state.  To describe this system we use the two-channel Hamiltonian
\begin{align}
\hat{H} &= \sum_{\vect{k}} \epsilon_{\vect{k},\text{Li}} \hat c^\dag_{\vect{k}}\hat c_{\vect{k}}+\sum_{\vect{k}} \epsilon_{\vect{k},\text{K}} \hat d^\dag_{\vect{k}}\hat d_{\vect{k}}
 + \sum_\vect{k} \left[\epsilon_{\vect{k},M}+\epsilon_M(B)\right]\hat b_\vect{k}^\dag \hat b_\vect{k} \nonumber\\ &+ \frac{g}{\sqrt{V}}\sum_{\vect{k},\vect{q}} \chi(\vect{k})
\left(\hat b^\dag_\vect{q} \hat c_{\vect{q}/2+\vect{k}}\hat d_{\vect{q}/2-\vect{k}} + \hat d^\dag_{\vect{q}/2-\vect{k}} \hat c^\dag_{\vect{q}/2+\vect{k}} \hat b_\vect{q} \right),
\label{eq:twochannel}
\end{align}
where the first line defines the non-interacting Hamiltonian
$\hat H_0$.  Here, $V$ is the total system volume,
$\hat c^\dagger_{\vect{k}}$ ($\hat c_{\vect{k}}$) creates
(annihilates) a Li fermion with momentum $\hbar\vect{k}$ and
single-particle energy
$\epsilon_{\vect{k},\text{Li}} = \frac{\hbar^2 k^2}{2m_\text{Li}}$,
and $\hat d^\dagger_{\vect{k}}$ ($\hat d_{\vect{k}}$) creates
(annihilates) a K impurity atom in the K$\ket{3}$ state with
dispersion
$\epsilon_{\vect{k},\text{K}} = \frac{\hbar^2 k^2}{2m_\text{K}}$,
where we define $k\equiv|\vect{k}|$.  The closed-channel molecule is
created (annihilated) by $\hat b^\dagger_{\vect{k}}$
($\hat b_{\vect{k}}$). It has the dispersion
$\epsilon_{\vect{k},M} = \frac{\hbar^2
  k^2}{2(m_\text{K}+m_\text{Li})}$,
and a bare energy relative to the scattering threshold,
$\epsilon_M(B)=\delta\mu (B-B_c)$.  Here $\delta\mu$ is the
differential magnetic moment between the open and closed channels, and
$B_c$ denotes the threshold crossing of the bare molecular state
\cite{Chin2010fri}.

Close to the Feshbach resonance, the scattering length $a$ diverges
and the interaction between the K impurities and the Li atoms is
predominantly mediated by exchange of the closed-channel molecule. We
therefore neglect the background scattering potential in the open
channel \cite{Naik2011fri}.  The strength of the coupling between the
open and closed channels is given by $g$, and we take a form factor
$\chi(\veck)=1/[1+(r_0k)^2]$, which accounts for the finite extent
$r_0$ of the closed-channel wave function $\sim e^{-r/r_0}/r$.

The parameters of the model $\delta\mu$, $B_c$, $g$, and $r_0$ are
fully determined by known experimental parameters. First, the
differential magnetic moment has recently been measured to be
$\delta\mu = h\times$2.35(2) MHz/G \cite{Cetina2015doi}. Second, close
to resonance, the scattering length may be parametrized as
\begin{align}
a=a_\text{bg}\left(1+\frac{\Delta B}{B_0-B}\right)\approx a_\text{bg}\frac{\Delta B}{B_0-B},
\label{eq:scat0}
\end{align}
where $B_0$ is the center of the Feshbach resonance with width
$\Delta B=0.880\,$G and background scattering length
$a_\text{bg}=63.0\,a_0$ \cite{Naik2011fri}. To connect with our model,
we consider the on-shell two-body scattering amplitude $f(k)$, which
for the Hamiltonian \eqref{eq:twochannel} is given
by~\cite{Schmidt2012epb}
\begin{align}\label{eq:scattamp}
  f(k) &= \frac{\mu_\text{red} g^2 \chi(\veck)^2}{2\pi\hbar^2}\left[-\frac{\hbar^2k^2}{2\mu_\text{red}}+\epsilon_M(B)-
\frac{g^2\mu_\text{red} }{4\pi\hbar^2 r_0[1-ikr_0]^2}\right]^{-1},
\end{align}
where
$\mu_\text{red}=m_\text{Li} m_\text{K}/(m_\text{Li}+ m_\text{K})$ is
the reduced mass and $\veck$ is the 
relative scattering wave vector. Using the low energy expansion
$f^{-1}(k)\approx-a^{-1}+\frac12r_\text{eff}k^2-ik$, with
$r_\text{eff}$ the effective range, we thus identify
\begin{align}
a& = \frac1{\frac1{2r_0}+2R^*\mu_\text{red}\delta\mu(B-B_c)/\hbar^2}, \label{eq:scat}\\
r_\text{eff}&=-2R^*+3r_0-4r_0^2/a,
\end{align}
where $R^*\equiv \hbar^4\pi/(\mu_\text{red}^2g^2)$ is the
range parameter of the Feshbach resonance
\cite{Bruun2004eto,Petrov2004tbp}. Comparing Eqs.~\eqref{eq:scat0} and
\eqref{eq:scat} yields
\begin{align}
  R^*&=\frac{\hbar^2}{2\mu_\text{red}a_\text{bg}
       \delta\mu\Delta B}, \label{eq:rstar} \\
  B_0-B_c&=\frac12\Delta B a_\text{bg}/r_0. \label{eq:bbres}
\end{align}
Equation~\eqref{eq:rstar} relates $R^*$, and thus the coupling
constant $g$, to the known experimental parameters.  The extent of the
closed-channel wave function $r_0$ in turn follows by comparing
Eq.~\eqref{eq:bbres} to the theoretical prediction from quantum defect
theory~\cite{Goral2004aao,Szymanska2005cco},
$B_0- B_c=a_{\rm{bg}}\Delta B/\bar{a}$, where
$\bar{a}=0.955 l_{\rm{vdw}}$ and $l_{\rm{vdw}}=40.8\,a_0$ is the van
der Waals length \cite{Naik2011fri}. Thus we obtain $r_0=\bar{a}/2$.
Finally, $B_0$ was obtained in Ref.~\cite{Cetina2015doi}, allowing the
determination of $B_c$.

\subsection{Truncated Basis Method}

To model a mobile impurity as in the experiment, we consider an
approximate wave function for the zero-momentum impurity that
incorporates the scattering of a single particle out of the Fermi sea:
\begin{align}
\ket{\psi_{\boldsymbol{\alpha}}} &= \alpha_{0} \hat d^\dag_{\mathbf{0}} \ket{\rm{FS}}
+ \sum_\vect{q} \alpha_\vect{q} \hat b^\dag_\vect{q} \hat c_{\vect{q}} \ket{\rm{FS}}
+  \sum_{\vect{k}, \vect{q}}
\alpha_{\vect{k}, \vect{q}}\hat d^\dag_{\vect{q}- \vect{k}}
\hat c^\dag_{\vect{k}} \hat c_{\vect{q}} \ket{\rm{FS}}.
\label{eq:tbm}
\end{align}
Here, the first term on the right hand side describes the product
state of the impurity K atom at zero momentum and the ground state of
the non-interacting Li Fermi sea
$\left| \rm{FS}\right> = \prod_{|\vect{k}|<k_F}\hat c^\dag_{\vect{k}}
\ket{0}$,
where $k_F$ is the Fermi momentum, which is related to the Fermi
energy by $\epsilon_F=\hbar^2 k_F^2/(2m_\text{Li})$.  The last two
terms correspond, respectively, to the impurity binding a Li atom to
form a closed-channel molecule, and the impurity exciting a particle
out of the Fermi sea, in both cases leaving a hole behind. When using
the TBM, we focus on zero temperature in order to capture the purely
quantum evolution of the impurity.  For convenience, within this model
we also take $r_0\to0$, which formally requires taking the bare
crossing $B_c\to\infty$ to keep $a$ finite. This approximation is
justified, as $R^*$ exceeds $r_0$ by about two orders of magnitude.

Truncated wave functions of the form \eqref{eq:tbm} have been used
extensively in the study of Fermi polarons in ultracold atomic gases,
starting with the work of Chevy \cite{Chevy2006upd}. While most of the
previous work has focused on equilibrium properties, recently it has
been proposed that these wave functions may be extended to dynamical
problems using a variational approach to obtain the equations of
motion \cite{Parish2013hpf}, for instance to calculate the decay rate
of excited states.

Here, we adapt the use of truncated wave functions for the Fermi
polaron to the calculation of the dynamical response of the impurity
to an interaction quench.  For a perfect quench and at zero
temperature, the quantity measured in experiment corresponds to the
overlap between the interacting and non-interacting states of the
system, i.e., we have~\cite{Goold2011oca,Knap2012tdi}
\begin{align}
S(t)=\left<\psi_0(t)|\psi_\text{int}(t)\right>=\left<\psi_0\right|e^{i\hat H_0t/\hbar }e^{-i\hat Ht/\hbar}\left|\psi_0\right> .
\label{eq:St}
\end{align}
Here $\ket{\psi_0}\equiv\hat d_{\vect{0}}^\dagger\ket{\rm{FS}}$ is the
initial non-interacting state of energy $E_0$, and $\psi_{\rm int}(t)$
is the state after a quench at time $t=0$ from zero to finite impurity
interactions with the Fermi sea. Formally expanding in a complete set
of states for the single impurity problem, the Ramsey signal
\eqref{eq:St} then becomes
\begin{align} \label{eq:rams}
S(t)=\sum_{j} \left|\left<\psi_0 | \phi_j \right>\right|^2 e^{-i(E_j-E_0)t/\hbar},
\end{align}
where $|\phi_j\rangle$ is an eigenstate of the interacting Hamiltonian
with energy $E_j$. However, this requires one to solve the entire
problem which is generally not possible for a mobile impurity. Thus,
within the Truncated Basis Method (TBM), we restrict the
Hilbert space to wave functions of the form \eqref{eq:tbm} and
diagonalize the Hamiltonian within this truncated basis.  As we shall
see, this truncation permits an extremely accurate description of the
initial quantum dynamics of the impurity.

For small $t$, we expand $e^{-i\hat Ht/\hbar}$ to find
\begin{equation}
  S(t) \approx1-(t/\tau_F)^2\frac{(1+m_\text{Li}/m_\text{K})^2}{3 \pi k_FR^*},
\label{eq:shortt}
\end{equation}
with $\tau_F$ the Fermi time.  This reveals that the short-time
dephasing dynamics of $S(t)$ is completely determined by the two-body
properties, which are captured exactly by the TBM.  As we will see
below, the TBM describes the impurity behavior also beyond the
two-body timescale since higher order correlations and multiple
particle-hole excitations take longer to build up. Indeed, for a
mobile impurity and for sufficiently weak attraction where the
attractive polaron is the ground state, the TBM correctly describes
the long-time behavior
$S(t) \to |\alpha_0|^2 e^{-i\varepsilon_p t/\hbar}$. Here,
$|\alpha_0|^2$ is the polaron residue (squared overlap with the
non-interacting state) and $\varepsilon_p$ is the polaron energy,
which are both accurately determined using a wave function of the
form~\eqref{eq:tbm}~\cite{Vlietinck2013qpo}.

With the TBM we consider zero temperature in order to isolate the
quantum dynamics of the impurity. To better model the experiment, in
principle one can extend the TBM to finite temperature by taking the
initial state to be a statistical thermal distribution involving
multiple impurity momenta. However, a more convenient approach at finite temperature is described in the next section.

\subsection{Functional Determinant Approach} \label{sec:fda}

At times $t$ substantially exceeding $\tau_F$, the full description of
the impurity dynamics requires the inclusion of multiple particle-hole
pair excitations as well as the effect of finite temperature, both of
which present a theoretical challenge.  In order to study and describe
both effects, we employ the Functional Determinant Approach (FDA)
\cite{Levitov1996ecs,Levitov1993cdi,Klich2003fcs,Knap2012tdi}.

In the FDA the impurity is treated as an infinitely heavy object. 
In this limit, the FDA provides an exact solution of the dynamical 
many-body problem at arbitrary temperatures and times.
The justification of the infinite mass approximation, which will be
discussed in more detail in Section \ref{SectionComparisonModels}, is
rooted in two observations.  First, in our experiment, the mass of the
K impurities is much larger than that of the Li atoms (mass ratio
$m_\text{K}/m_\text{Li}\approx 6.7$) which constitute the surrounding
Fermi gas. Therefore, the recoil energy gained by the K impurities due
to the scattering with a Li atom is small. We estimate the typical
recoil momentum $k_R$ by averaging over all possible scattering
processes on the Fermi surface, yielding $k_R = 4k_F/3$. From that we
obtain an estimate for the typical recoil energy
$E_R = \frac{16}{9}\frac{m_\text{Li}}{m_\text{K}} \epsilon_F \approx
0.25 \epsilon_F$,
which determines a typical time scale
$\tau_R=\hbar/E_R\approx 4\tau_F$, up to which one expects recoil to
have a minimal effect on the many-body quantum dynamics,
cf.~Section~\ref{SectionRecoilEffect}.  Second, at times exceeding the
thermal time scale $\tau_T=\hbar/(k_BT)$, which in our experiment is
given by $\tau_T\approx 6\tau_F$, thermal effects due to the averaging
over various statistical realizations become relevant. The resulting
thermal fluctuations disrupt the coherent quantum propagation of the
impurity, and hence, for times $t>\tau_T$, mask the effect of recoil
\cite{Rosch1999qct}.

To a good approximation, we may thus take the limit of infinite
impurity mass, which admits the mapping of Eq.~\eqref{eq:twochannel}
onto the bilinear Hamiltonian
\begin{align}
\hat H = \epsilon_M(B) \hat m^\dag\hat m + \sum_{\vect{k}} \epsilon_{\vect{k}} \hat c^\dag_{\vect{k}}\hat c_{\vect{k}}
+ g \sum_\veck \chi(\veck) [\hat m^\dag \hat c_{\veck} + \hat m \hat c^\dag_{\veck}].
\label{eq:h}
\end{align}
Here, $\hat m^\dagger$ is the creation operator of the localized
closed channel molecule and the interaction is described by the
annihilation of a Li atom converting the empty impurity molecular
state into an occupied one.  By taking the limit $m_\text{K}\to\infty$
we obtain a modified reduced mass $\mu'_\text{red}=m_\text{Li}$, which
differs by a factor of $40/46$ from the experimental one. This needs
to be taken into account when identifying the microscopic
parameters. To ensure, in particular, that the off-diagonal coupling
$g$ in \eq{eq:h} remains of the same strength as in the experiment, a
reduced resonance parameter $R'^*=(40/46)^2 R^*$ has been used, which
we do for all data shown in the main text. Using these
identifications, the model Eq.~\eqref{eq:h} also accurately describes
the short-time dynamics as given by Eq.~\eqref{eq:shortt}, cf. Fig.~2
in the main text.

The calculation of time-resolved, many-body expectation values such as
Eq.~\eqref{eq:St} at arbitrary temperature presents a theoretical
challenge. However, for the model \eqref{eq:h}, we are able to
calculate the time-resolved Ramsey response in an exact way using the
FDA~\cite{Klich2003fcs,Knap2012tdi}. This is based on the observation
that for bilinear Hamiltonians thermal expectation values in the
many-body Fock space can be reduced to determinants in the
single-particle space by virtue of the identity
\begin{align}
 \tr [\hat \rho \ e^{\hat Y_1}e^{\hat Y_2}\ldots ] = \det [1-\hat n+\hat n\ e^{\hat y_1}e^{\hat y_2}\ldots].
 \label{eq:fda}
\end{align}
Here $\hat Y_1, \hat Y_2, \ldots$ are many-body operators,
$\hat y_1, \hat y_2, \ldots$ are their single-particle counterparts,
$\hat \rho$ is the many-body density matrix describing the state of
the system, and $\hat n = 1/[e^{\beta( \hat h_0-\mu)}+1]$ is the
occupation operator defined in the single-particle space, with $\mu$
the fermion chemical potential. A specific example for \eq{eq:fda} is
the perfect quench Ramsey response, which at finite temperature is
given by~\cite{Knap2012tdi}
\begin{align}\label{FDAS}
 S(t)= \tr [\hat \rho \ e^{i\hat H_0 t}e^{-i \hat H t}] = \det [1-\hat n+\hat n\ e^{i\hat h_0 t}e^{-i\hat h t}].
\end{align}
Here,
$\hat H_0=\sum_{\vect{k}} \epsilon_{\vect{k}} \hat
c^\dag_{\vect{k}}\hat c_{\vect{k}}$
is the free Hamiltonian of the Li Fermi gas and $\hat H$ is the
Hamiltonian in the presence of impurity scattering given in \eq{eq:h},
while $\hat h_0$ and $\hat h$ are their single-particle
counterparts. A numerical evaluation of Eq.~\eqref{FDAS} then only
requires a calculation of the single particle orbitals and energies in
order to obtain the single-particle determinant.

\section{Role of physical processes on different time scales}\label{SectionComparisonModels}
The combination of both our theoretical approaches allows us to
accurately model the physics at various time scales in our
experiment. Making use of the fact that the FDA and the TBM differ
distinctly in their treatment of multiple particle-hole excitations,
the impurity mass, and finite temperature, we can use a comparison of
their predictions to determine the role of these processes and effects
in the many-body non-equilibrium dynamics of our experiment.  To keep
the analysis transparent, in this section we still assume that a
perfect quench is performed.

\subsection{Multiple particle-hole excitations}\label{multiplephfluctuations}

In order to analyze the role of multiple particle-hole excitations, we
first consider the limit of a fixed (infinitely heavy) impurity at
zero temperature.  In this scenario, the FDA yields the exact solution
of the impurity problem. Since, in this case, the TBM only differs
from the FDA by its neglect of multiple particle-hole excitations, a
comparison of the predictions of the two methods allows us to isolate
the effect of these excitations.

\begin{figure*}[tbh]
        \centering
        \includegraphics[width=16cm]{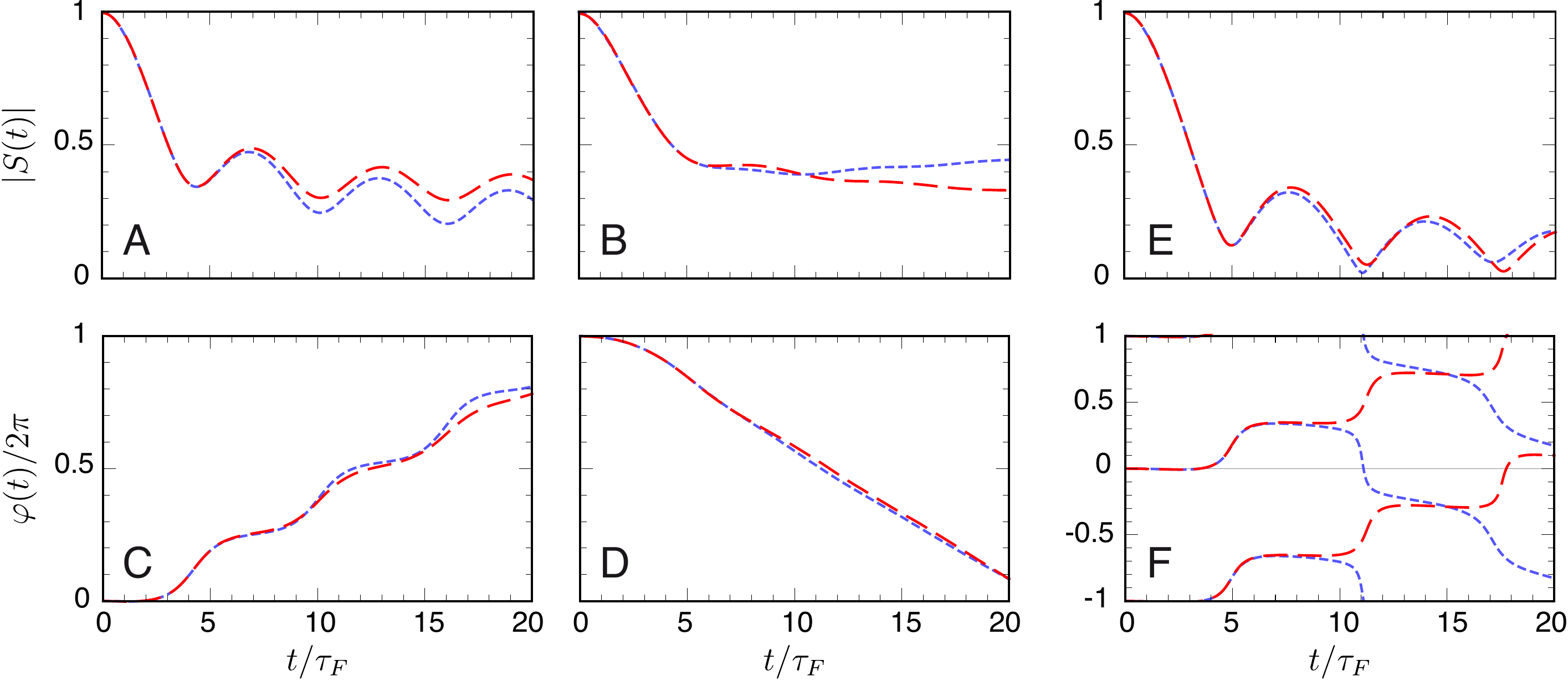}
        \caption{ \textbf{Effect of multiple particle-hole
            fluctuations}. Taking the idealizing limit of zero
          temperature and infinite impurity mass, we compare the
          Ramsey response for a perfect quench (top: amplitude,
          bottom: phase) obtained exactly with the FDA (red, long
          dashed) to the one obtained with TBM (blue, short dashed)
          for \fc{{\bf A}, {\bf C}} $X=-0.23$, \fc{{\bf B}, {\bf D}}
          $X=0.86$, and \fc{{\bf E}, {\bf F}} $X=0.08$. For this
          comparison, we take $r_0=0$ and $k_F R'^*=1.1(40/46)^2$.  }
        \label{fig:TBMvsFDA} 
\end{figure*}

In \fig{fig:TBMvsFDA} we display the predictions for the Ramsey
response using the two theoretical approaches.  We find that both
theoretical predictions agree extremely well at short times. In
particular, for both the amplitude and phase of $S(t)$, our results
imply that multiple particle-hole excitations start to influence our
observables at a time scale of around $6\tau_F$, and only become
prominent beyond $10\tau_F$. Thus, at shorter time scales, multiple
particle-hole excitations can be neglected when predicting the results
of the Ramsey measurements.

We note that the fixed impurity scenario is a worst-case scenario for
the TBM: At $T=0$, the infinitely heavy impurity is subject to the
orthogonality catastrophe with an associated power-law decay of the
Ramsey contrast at long times \cite{Anderson1967ici}. This decay,
which arises due to an infinite number of particle-hole fluctuations
and which leads to a vanishing quasiparticle weight, is exactly
incorporated in the FDA. By contrast, in the long-time limit, the TBM
predicts the saturation of $|S(t)|$ to a constant value (see
\fig{fig:TBMvsFDA}), corresponding to a spurious finite residue.
However, for a \textit{mobile} impurity at zero temperature, recoil
becomes relevant.  These recoil effects lead to the absence of the
orthogonality catastrophe \cite{Rosch1999qct}, and thus to an
increased accuracy of the TBM in the case of finite impurity mass.

Generally, one expects that the relevant time scale for multiple
particle-hole excitations is closely related to the Fermi time
$\tau_F$. As discussed above, we find that such excitations become
relevant for a description of $S(t)$ only at around $6\tau_F$ or
beyond. This observation can be understood in a twofold way.  First,
in the equilibrium case it was found that contact interactions in the
Fermi polaron problem lead to an approximate cancellation of terms
involving identical fermions, thus suppressing the emergence of
multiple particle-hole fluctuations~\cite{Combescot2008nso}. Our
observation may hence be interpreted as a generalization of these
findings to the non-equilibrium case.  Second, the spectrum of the
Fermi polaron problem features a dominant contribution involving the
excitation of fermions from the bottom of the Fermi sea to the Fermi
surface \cite{Knap2012tdi}. As discussed in Ref.~\cite{Knap2012tdi},
these excitations manifest themselves as oscillations with period
$2\pi\tau_F$ in the Ramsey contrast $|S(t)|$. Such a bottom of the
band excitation is also present in the truncated wavefunction
\eqref{eq:tbm}, and indeed the remarkable agreement of the TBM with
the exact solution from the FDA up to the time $2\pi\tau_F$ suggests
that this effect can be captured by single-particle hole excitations.

\subsection{Impurity mass}\label{SectionRecoilEffect}

\begin{figure*}[tbh]
        \centering        
        \includegraphics[width=16cm]{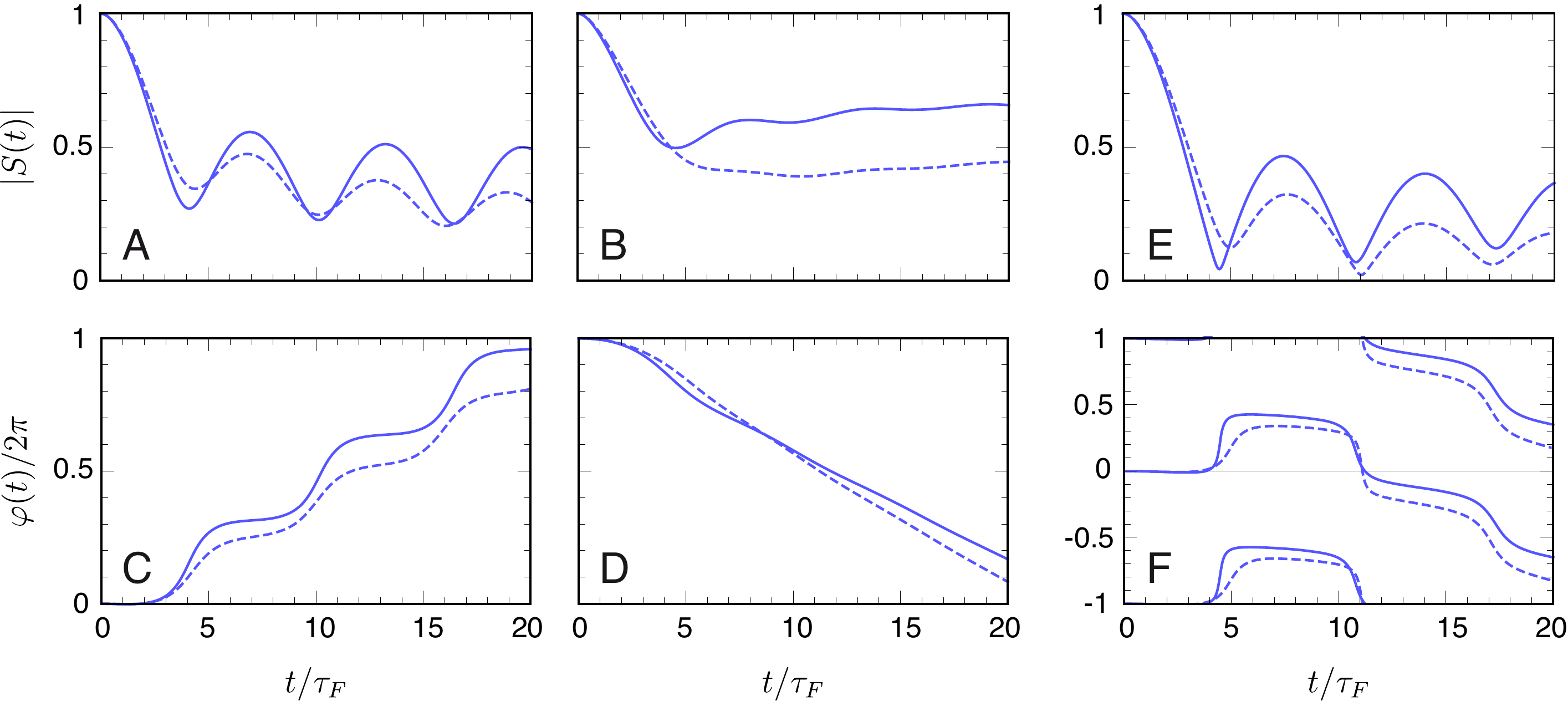}
        \caption{\textbf{Effect of the impurity motion on the
            short-time dynamics.}  Amplitude (top) and phase (bottom)
          of the perfect quench zero temperature Ramsey response
          $S(t)$ as a function of time for \fc{{\bf A}, {\bf C}}
          $X=-0.23$, \fc{{\bf B}, {\bf D}} $X=0.86$, and \fc{{\bf E},
            {\bf F}} $X=0.08$. We compare the results of the TBM
          obtained for $m_\text{K}=(40/6)m_\text{Li}$ and
          $k_F R^*=1.1$ (solid) with the TBM results for fixed
          impurities $m_\text{K}\to\infty$ and $k_F R'^*=1.1(40/46)^2$
          (dashed).  }
        \label{fig:mKvsinf} 
\end{figure*}

As discussed in the main text, our experimental findings are well
described by the static impurity approximation, although the impurity
has finite mass.  To quantify the effect of the finite impurity mass,
we study here the case of zero temperature. This allows us to isolate
the effect of finite impurity recoil from the influence of thermal
fluctuations, which will become dominant beyond times
$\tau_T\approx 6\tau_F$, as discussed in the section below.  In order
to estimate at which time scale recoil becomes important, we make use
of the capability of the TBM to describe impurities of arbitrary
mass. Furthermore, our analysis in Sec.~\ref{multiplephfluctuations}
shows that the TBM yields highly accurate results for the short-time
dynamics of $S(t)$.  Accordingly, in \fig{fig:mKvsinf} we display the
Ramsey response for a static impurity and for the experimentally
relevant impurity mass, both calculated within the TBM. We see that
for both amplitude and phase, the impurity motion only results in a
small difference in the Ramsey signal at times $t \lesssim 4 \tau_F$.
Physically, this time scale corresponds to the effective recoil time
$\tau_R$ associated with Li collisions on K atoms, which we estimated
in Sec.~\ref{sec:fda} to be $\tau_R\approx4\tau_F$, in agreement with
our findings here.  At times exceeding $\tau_R$, we find that the
dynamics is indeed affected by the finite impurity mass.  However, at
such times, thermal fluctuations dominate the behavior in experiment,
as we now discuss.

\subsection{Temperature}\label{SectionTemperatureEffect}

\begin{figure*}[tbh]
        \centering
        \includegraphics[width=16cm]{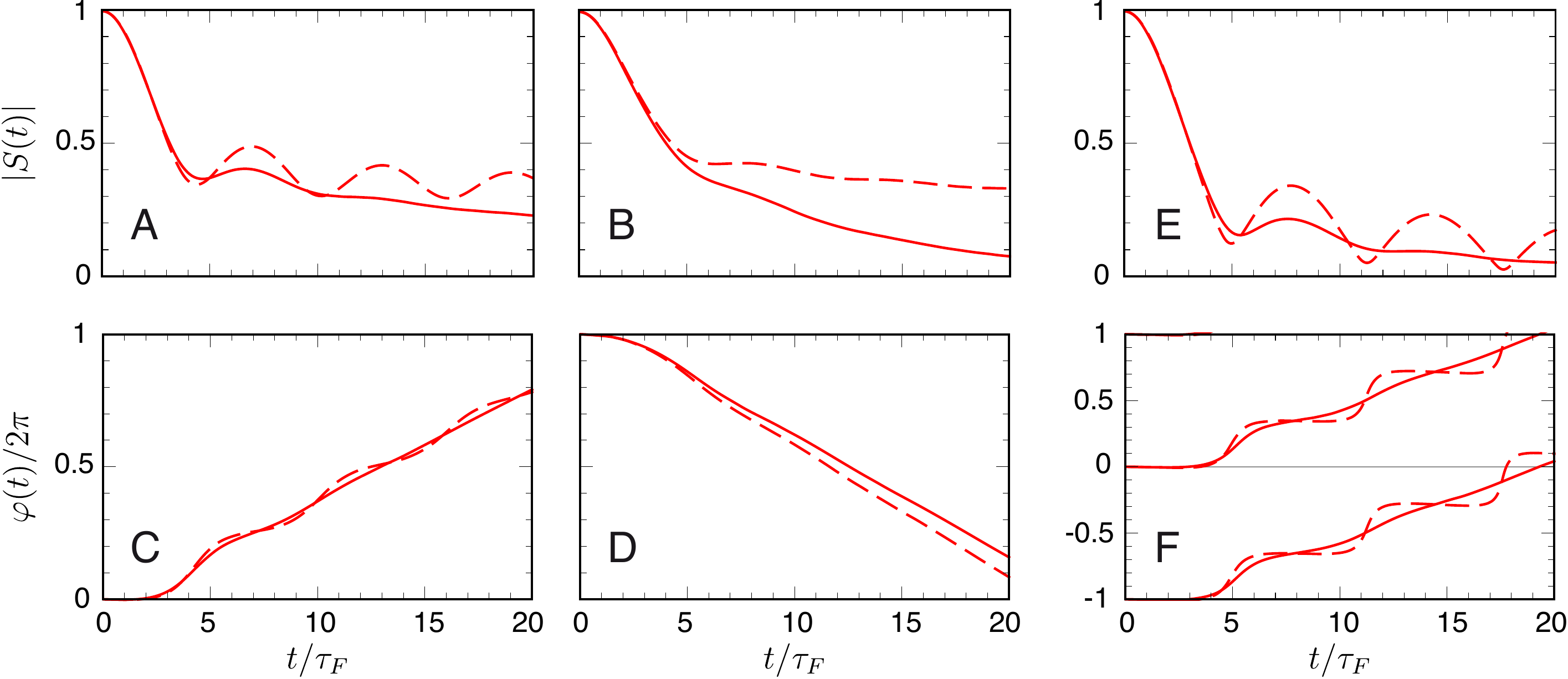}
        \caption{\textbf{ Effect of finite temperature on the impurity
            dynamics}. We compare the Ramsey signal (upper panels:
          amplitude, lower panels: phase) for an infinitely heavy
          impurity obtained from an exact FDA calculation at zero
          (long dashed) and finite temperature (solid curves). The
          ordering of the graphs is as in the main text: \fc{{\bf A},
            {\bf C}} $X=-0.23$, $T/T_F=0.17$, \fc{{\bf B}, {\bf D}}
          $X=0.86$, $T/T_F=0.16$, and \fc{{\bf E}, {\bf F}} $X=0.08$,
          $T/T_F=0.18$. We assume a perfect quench and choose $r_0=0$
          as well as $k_F R^*=1.1$, i.e., $k_F R'^*=1.1(40/46)^2$.  }
        \label{fig:TempDep} 
\end{figure*}

At long times, the time evolution reduces to a simple exponential
decoherence of $S(t)$. The time scale at which this crossover to
exponential decay takes place is given by the thermal time scale
$\tau_T$. In our experiment, where $T/T_F\approx0.15$, this
corresponds to $\tau_T\approx 6\tau_F$ and, hence, we observe both
regimes within the dynamical range probed in our experiment.

In this section, we use finite-temperature FDA calculations to gauge
the role of temperature in the impurity dynamics.  To this end we
compare the results for the Ramsey signal at zero and finite
temperature for the experimentally realized parameters. The results
are shown in Fig.~\ref{fig:TempDep}. We indeed find that at times
$\sim 6 \tau_F$ the time evolution at finite temperature starts to
deviate from the purely quantum behavior. Finite temperature
leads to an exponential decoherence of the Ramsey signal
and has the consequence that thermal
fluctuations dominate over the impurity motion at times
$t \gtrsim6 \tau_F$ \cite{Rosch1999qct}. Hence they mask the
effect of impurity recoil as discussed in Sec.~\ref{sec:fda}.

Overall, the conditions in our experiment give rise to three competing
time scales. Multiple particle hole excitations become relevant for
our measurement of $S(t)$ at around $6\tau_F$, the recoil time is
$\tau_R\approx4\tau_F$, and the thermal scale is set by
$\tau_T\approx6\tau_F$. A comparison of these scales reveals the
reason for the remarkable agreement between the FDA and experiment:
Recoil is only weakly probed at short times $t<\tau_R$, while its
effect is washed out by the thermal fluctuations at long times
$t > \tau_T \approx \tau_R$.

\section{Role of interaction during finite-length rf pulses \label{sec:S3}}

In this section, we analyze the role of the `imperfect' interaction
quench in our experiments, where residual interactions are present
during the rf pulses. Furthermore, we discuss how our findings pave
the way towards the use of our experimental techniques to exert
control over many-body states in real time.

\subsection{Idealized versus realized Ramsey scenario}\label{sec:S3A}

\begin{figure*}[tbh]
\centering
 \includegraphics[width=14cm]{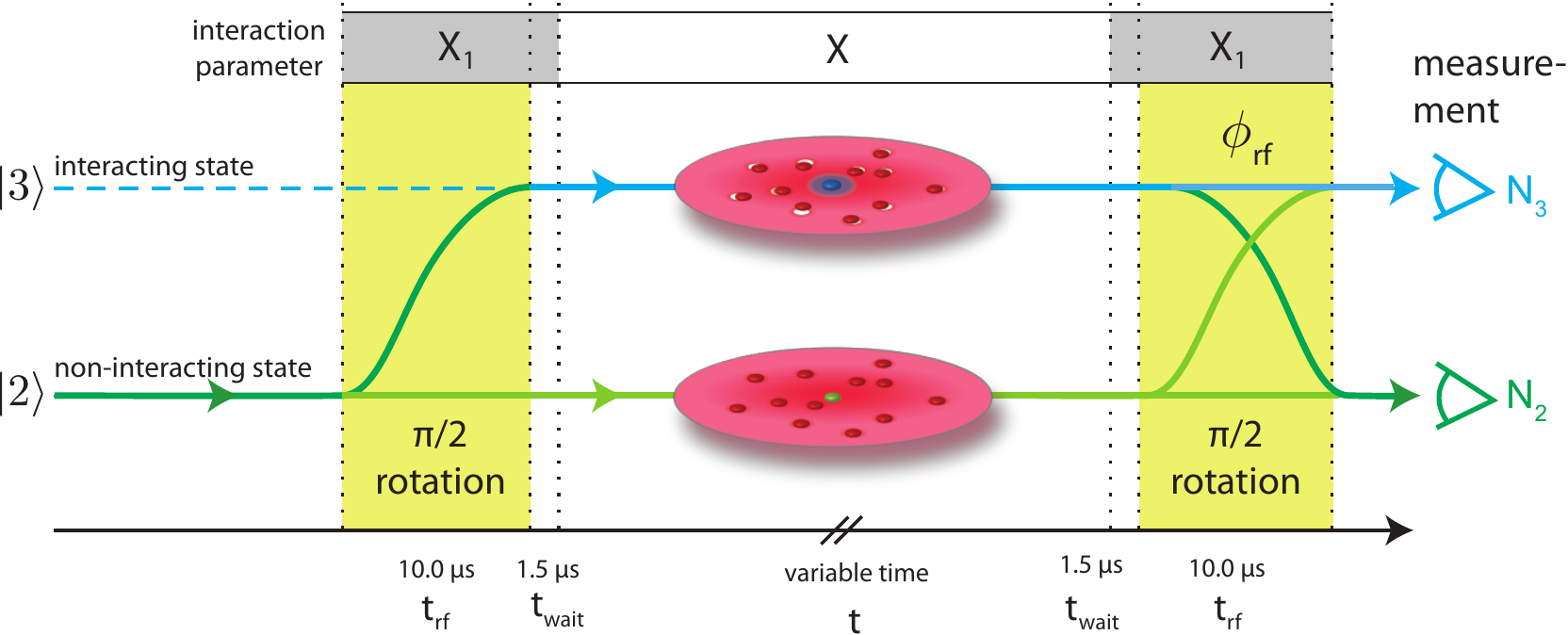}
 \caption{ \textbf{Schematic of the experimental Ramsey procedure.}
   The K atoms start out in the hyperfine state K$\ket{2}$, which is
   effectively non-interacting 
   with the Fermi sea. A 10 $\mu$s (3.4
   $\tau_F$) long square $\pi/2$ pulse is applied in the presence of
   weak interactions between the K$\ket{3}$ atoms and the Li atoms,
   quantified by the interaction parameter $X_1$.  We then use optical
   control of our Feshbach resonance to rapidly (in less than 200 ns 
   (0.08 $\tau_F$)) quench the system into the strongly interacting
   regime (interaction parameter $X$). After a variable interaction
   time $t$\, we optically shift the interaction strength back to $X_1$, and
   then close the Ramsey sequence by a second $\pi/2$ pulse.
   We vary the phase of this pulse by shifting the phase of the rf 
   source by $\phi_{\rm{rf}}$ before the second pulse is applied.
}
\label{fig:rfsequence}
\end{figure*}

Thus far, we have assumed the idealized scenario of a perfect
two-pulse Ramsey scheme.  In this case, the initial spin state of the
impurity (K$\ket{2}$ in the experiment) is non-interacting with the Li
Fermi sea and there are no interactions during the applied rf $\pi/2$
pulses.
Each pulse then yields a perfect rotation on the Bloch sphere,
e.g., the initial state K$\ket{2}$ is transformed into the spin-state
superposition $(\text{K$\ket{2}$}+\text{K$\ket{3}$})/\sqrt{2}$.  For
such a perfect Ramsey sequence, the measured Ramsey signal $S(t)$
gives the overlap between the time-evolved interacting and
non-interacting states of the system \cite{Goold2011oca,Knap2012tdi},
yielding Eqs.~\eqref{eq:St} and \eqref{FDAS} for zero and finite
temperature, respectively.  In this idealized scenario, the Fourier
transform of $S(t)$ corresponds to the excitation spectrum of the system
in linear response~\cite{Mahan1990mpp},
\begin{align}\label{ASRelation}
A(\omega) =  \re \int^\infty_0 \frac{dt}{\pi} e^{i\omega t} S(t),
\end{align}
where $\omega$ is the frequency of the applied field.

In our experiments, however, residual interactions are present during
the $\pi/2$ pulses, which take a finite time to be completed. As shown in the illustration of our experimental
sequence in Fig.~\ref{fig:rfsequence}, the state K$\ket{3}$ can
already interact with the Li cloud during the $\pi/2$ rotation, which
potentially affects the observed dynamics of the system. Specifically,
this stage of the experiment is performed at a detuning from the
Feshbach resonance which corresponds to a weak interaction strength
$X_1$ between the impurities and the Fermi sea (cf. Section
\ref{sec:rfpulse} and ~Fig.~\ref{fig:rfsequence}).  After preparing
the superposition state of the impurity spin, we quench the system to
strong interactions (interaction parameter $X$) by optically shifting
the Feshbach resonance \cite{Cetina2015doi}.  We previously focussed
on the complex non-equilibrium dynamics resulting from the strong
interactions $X$ during the time $t$.  In the following, we analyze
the effect of the residual interaction $X_1$ during the finite-duration $\pi/2$ spin
rotations. In particular, we investigate the impact of these weak
interactions during the rf pulses on the Ramsey response $S(t)$ and
the spectrum $A(\omega)$ as obtained from the Fourier transform
Eq.~\eqref{ASRelation}.

\subsection{Modelling of rf pulses within TBM} \label{sec:rfpulse}

In this section, we extend our modelling of the zero-temperature
impurity dynamics within the TBM to directly simulate the entire
experimental procedure, as illustrated in \fig{fig:rfsequence}. In order to
model the rf pulses, we explicitly include both K$\ket{2}$ and
K$\ket{3}$ spin states, as well as the rf field. This modifies the
Hamiltonian, Eq.~\eqref{eq:twochannel}, to
$\hat {\cal H}=\hat H +\hat H_{\rm rf}$ with the additional term
\begin{align}\label{rfHamil} 
\hat{H}_{\rm rf} &=  \frac{\Omega}{2i} \sum_{\vect{k}} \left(e^{i\phi_{\rm rf}} \hat{d}^\dag_{\vect{k}, 2} \hat{d}_{\vect{k},3} - e^{-i\phi_{\rm rf}} \hat{d}^\dag_{\vect{k},3} \hat{d}_{\vect{k},2}\right)
+ \sum_{\vect{k}}  (\epsilon_{\vect{k},\rm{K}} + \hbar(\omega_\text{rf}-\omega_0))
\hat{d}^\dag_{\vect{k},2}\hat{d}_{\vect{k},2}. 
\end{align}
Here, we have used the rotating wave approximation.  $\Omega$
corresponds to the strength of the rf field, $\phi_{\rm rf}$ is the
variable phase of the second rf pulse, and
$\hat{d}^\dag_{\vect{k},\sigma}$ creates a particle in the state
K$\ket{\sigma}$ with momentum $\hbar\vect{k}$.  Note that
$\hat{d}^\dag_{\vect{k}}\equiv\hat{d}^\dag_{\vect{k},3}$ in the
original two-channel Hamiltonian \eqref{eq:twochannel}.  The
interactions during the rf pulses cause a shift in the transition
frequency between the K$\ket{2}$ and K$\ket{3}$ states from the bare
transition frequency $\omega_0$ to $\omega_0+\varepsilon_{1}/\hbar$,
where $\varepsilon_{1}$ is the polaron energy at interaction parameter
$X_1$.  As described in Sec.~S5.B, we account for this shift by
adjusting the frequency of our rf pulses to
$\omega_\text{rf}= \omega_0+\varepsilon_{1}/\hbar$.

According to the last term in Eq.~\eqref{rfHamil}, the shift in the
frequency of the rf source from $\omega_0$ to $\omega_\text{rf}$
causes the observed signal to accumulate an additional phase
$(\omega_\text{rf} - \omega_0)t$ during the interaction time $t$. To
account for this, we introduce the phase
$\phi = \phi_\text{rf} + (\omega_\text{rf} - \omega_0)t$. We then
determine $|S(t)|$ and the phase $\varphi(t)$ by noting that the
Ramsey signal $(N_3 - N_2)/(N_3 + N_2)$ corresponds to a sine-wave
function of $\phi$ plus an offset, i.e., it takes the form
$F(t) + |S(t)| \cos (\phi - \varphi(t))$ with $F(t)$ a real,
$\phi$-independent function. This mirrors the experimental procedure,
where $F(t)$, $|S(t)|$, and $\varphi(t)$ appear as fit-parameters for
the Ramsey signal, see Sec.~S5.B.

\begin{figure*}[thb]
        \centering
        \includegraphics[width=16cm]{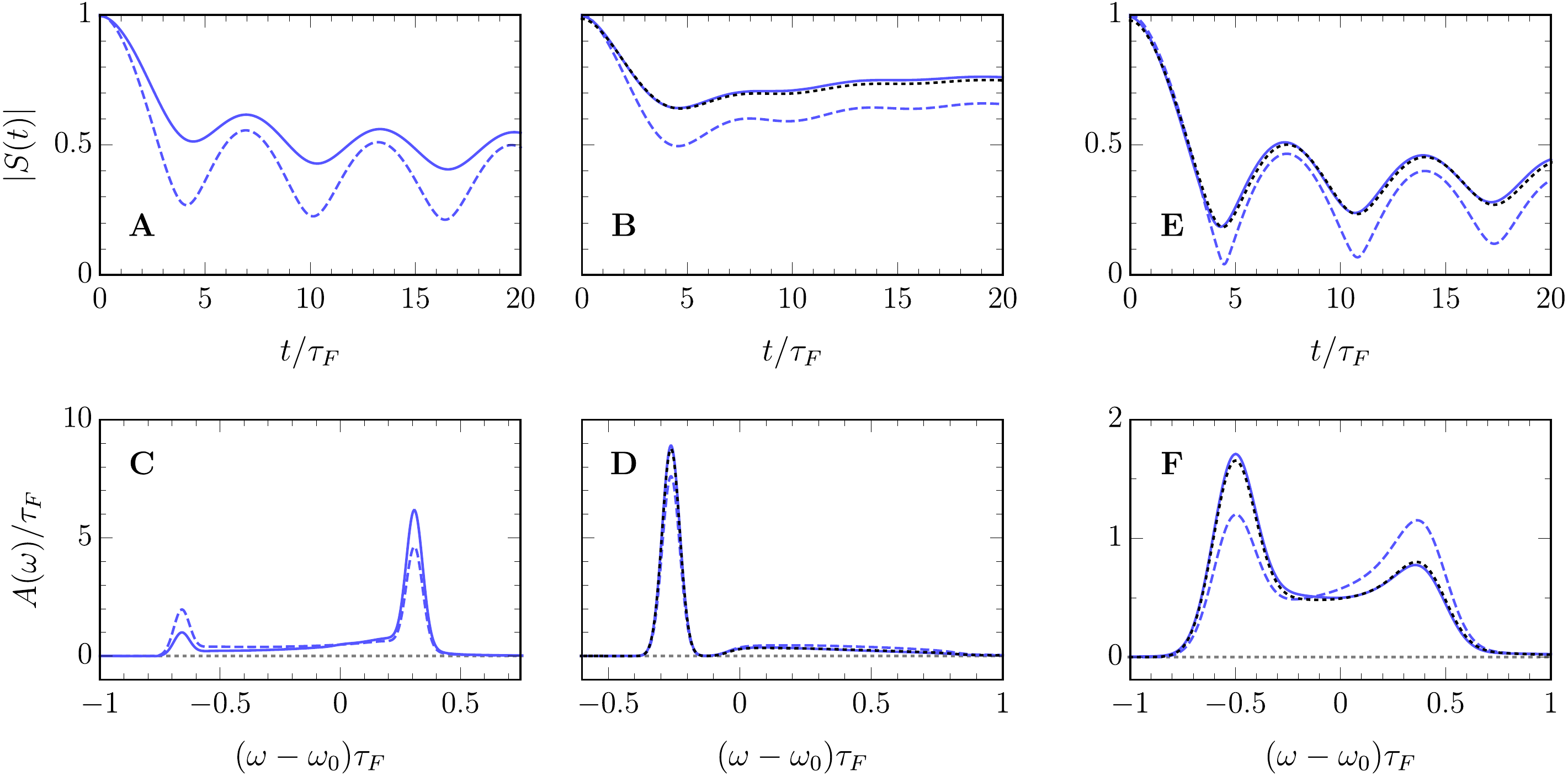}
        \caption{\textbf{Role of the residual interactions within
            TBM.}  We present the zero-temperature response $S(t)$ and
          the corresponding spectrum $A(\omega)$ for the
          perfect quench (dashed blue) and the actual experimental
          sequence shown in Fig.~\ref{fig:rfsequence} (solid blue). As
          in the main text, we take $k_F R^* = 1.1$ and the
          interaction parameters: \fc{{\bf A},
            {\bf C}} $X=-0.23$, $X_1=-3.9$, \fc{{\bf B}, {\bf D}} $X=0.86$, $X_1=5.8$, and
          \fc{{\bf E}, {\bf F}} $X=0.08$, $X_1=4.8$.
          For comparison, in ({\bf B}, {\bf D}, {\bf E}, {\bf F}), 
          we represent by black dotted lines the scenario where the
          initial state before the quench is approximated as a weakly
          attractive polaron --- see Sec.~\ref{sec:rfpulse} for details.  The
          spectra have been convolved with the experimental
          Fourier-limited rf spectral lineshapes, which are
          Gaussian-shaped with width $\sigma$, where
          $\sigma\tau_F=0.03$ for $X=0.86$, $-0.23$, and
          $\sigma\tau_F=0.1$ for $X=0.08$.  }
        \label{fig:imperfect} 
\end{figure*} 

Within the TBM, we determine the approximate eigenstates and
eigenvalues of $\hat{\cal{H}}$ within the more general class of
truncated wavefunctions:
\begin{align}\notag
  \ket{\psi_{\rm{rf}}} &= \left(\alpha_{0,3} \hat d^\dag_{0,3} +  \alpha_{0,2} \hat d^\dag_{0,2} \right) \ket{\rm{FS}}
  + \sum_\vect{q} \alpha_\vect{q} \hat b^\dag_\vect{q} \hat c_{\vect{q}} \ket{\rm{FS}}
  + \sum_{\vect{k} \vect{q}}
  \left(\alpha_{\vect{k} \vect{q},3}\hat d^\dag_{\vect{q}- \vect{k} \downarrow}
  \hat c^\dag_{\vect{k}} \hat c_{\vect{q}} 
  + \alpha_{\vect{k} \vect{q},2}\hat d^\dag_{\vect{q}- \vect{k},2}
  \hat c^\dag_{\vect{k}} \hat c_{\vect{q}} \right) \ket{\rm{FS}}.
\end{align}
To model the experimental quench sequence illustrated in
Fig.~\ref{fig:rfsequence}, we apply a series of time evolution
operators to the initial state consisting of a K$\ket{2}$ atom and the
Li Fermi sea. At the end of the sequence we then extract the number of
K atoms in states K$\ket{2}$ and K$\ket{3}$, respectively. We include
explicitly the rf pulses, the wait times, and the interaction time $t$
during which the system is strongly interacting. The results of this 
procedure are displayed in Fig.~2 of the main text.
Here, we account for slight additional experimental decoherence by
scaling the prediction for $|S(t)|$ as described in Section S.5C. 

In the upper panels of Fig.~\ref{fig:imperfect} we compare the Ramsey
response obtained by simulating the actual experimental sequence
(solid line) with that of the perfect quench scenario (dashed line).
We see that the residual interactions $X_1$ in experiment can indeed
influence the quantum evolution of the impurity. The difference in the
responses can be straightforwardly explained by assuming that the main
effect of $X_1$ is to produce a weakly interacting initial state.
Specifically, for weak attractive interactions $X_1>0$, the Ramsey
response can be approximated as
\begin{align}\label{eq:pol_quench}
  S(t)  \simeq Z \left<\psi_{X_1}\right|e^{-i\hat Ht/\hbar}\left|\psi_{X_1}\right>,
\end{align}
where $\left|\psi_{X_1}\right>$ is the ground state of the Hamiltonian
\eqref{eq:twochannel} at interaction parameter $X_1$, and $Z$ is the
corresponding polaron residue. Note that we cannot formally construct
a similar expression for the repulsive case $X_1<0$, since the
repulsive polaron is a metastable state, involving multiple
eigenstates of the Hamiltonian.
 
Referring to \fig{fig:imperfect}, the excellent agreement between the
approximation \eqref{eq:pol_quench} and the full Ramsey signal
provides strong evidence that the residual interactions $X_1$ produce
a weakly attractive initial state.  This is further supported by the
spectrum $A(\omega)$ shown in the bottom panels, where we see that the
residual interactions enhance the attractive polaron peaks for
$X=0.08$ and $0.86$.  A similar enhancement of the repulsive polaron
peak is observed for $X=-0.23$.  Hence we conclude that the explicit
modelling of the impurity dynamics using the full Hamiltonian
$\hat{\cal H}=\hat H+\hat H_\text{rf}$ is not essential for the
description of the dynamics during the initial $\pi/2$ spin rotation
and instead one can fully describe the time evolution using the
Hamiltonian \eqref{eq:twochannel}.

\subsection{Modelling of experimental procedure at finite temperature within FDA} 

\begin{figure*}[thb]
        \centering
        \includegraphics[width=16cm]{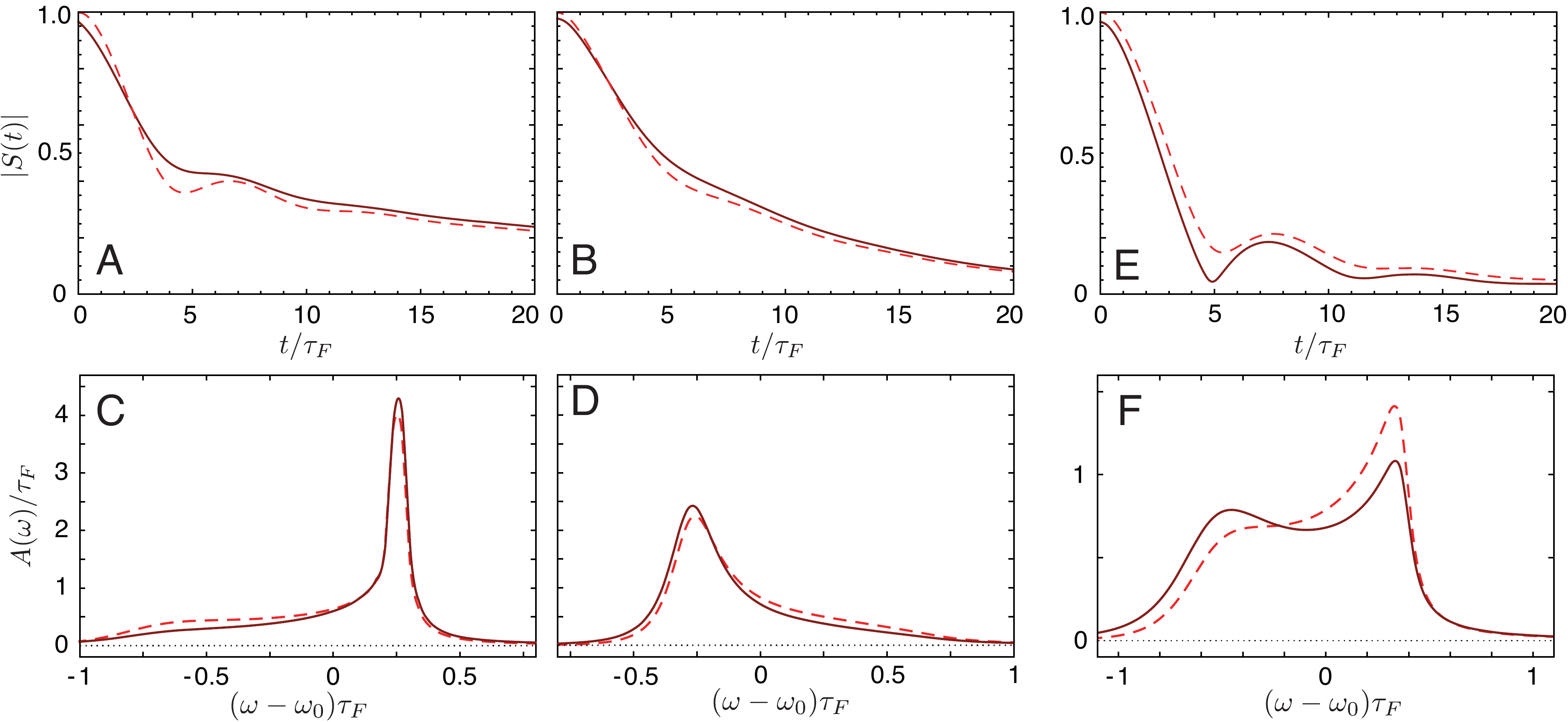}
        \caption{\textbf{Role of the residual interactions in the
            Ramsey sequence at finite temperature}. Upper panels: we
          compare the perfect quench Ramsey response (dashed) with a
          simulation of the experimental sequence (solid). Lower
          panels: we compare the linear-response excitation spectrum (dashed)
          with the Fourier transform of the signal obtained using the
          experimental sequence (solid). As in the main text, we take
          $k_F R^* = 1.1$ and the interaction parameters: \fc{{\bf A},
            {\bf C}} $X=-0.23$, $X_1=-3.9$, \fc{{\bf B}, {\bf D}} $X=0.86$, $X_1=5.8$, and
          \fc{{\bf E}, {\bf F}} $X=0.08$, $X_1=4.8$. The temperatures are
          $T/T_F=0.166,\,0.158,\,0.177$, respectively.  }
        \label{figtwostep} 
\end{figure*} 

The interplay between the residual interactions and finite 
temperature presents a further theoretical challenge.  In the
following, we use the FDA to simulate the experimental protocol
(Fig.~\ref{fig:rfsequence}) at finite temperature. To achieve this, we
exploit the finding from Sec.~\ref{sec:rfpulse} that the 
detailed dynamics of the rf-driven oscillations between the K$\ket{2}$ 
and K$\ket{3}$ states can be ignored when calculating $S(t)$.
Thus, we assume that the initial $\pi/2$ rotation effectively produces
a spin superposition $(\text{K}\ket{2}+\text{K}\ket{3})/\sqrt{2}$,
independently of the residual interaction $X_1$ of the impurity in the
state K$\ket{3}$ with the Fermi sea.  To account for the
dynamics due to the weak interaction $X_1$, we then let the
system evolve under this interaction for a hold time
$t_\text{h}=t_\text{rf}/2+t_\text{wait}$, which
models the dynamics at weak interaction $X_1$ 
as the result of a sudden switch-on of this interaction at the midpoint of the $\pi/2$ pulses.
After the hold time
$t_\text{h}$, the final quench to the strong interactions $X$ is
performed. For the measurement of the Ramsey contrast, this sequence is
reversed. Theoretically, this yields the modified time-dependent
overlap
\begin{equation}\label{modS}
S(t)=\tr \left[\hat\rho\, e^{i\hat H_0 (2t_\text{h}+t)}e^{-i\hat H_{1} t_\text{h}}e^{-i\hat H_{X} t}e^{-i\hat H_1 t_\text{h}}\right],
\end{equation}
where $\hat H_1$ and $\hat H_X$ denote the Hamiltonian
\eqref{eq:twochannel} at interaction strength $X_1$ and $X$,
respectively. Using the FDA, the expression Eq.~\eqref{modS} is
evaluated exactly according to Eq.~\eqref{eq:fda} at the experimental
temperature. As can be inferred from Eq.~\eqref{modS}, this simplified
model of the experimental protocol corresponds to a sequence of
interaction quenches.

In the upper panel of Fig.~\ref{figtwostep} we compare the result for
$|S(t)|$ at the experimental temperatures obtained for the
experimental sequence (solid lines) to the result for an idealized,
i.e., perfect quench, Ramsey sequence (dashed lines).  Similarly to
the case of zero temperature, we see that the time evolution at
$X_1$ has an experimentally observable effect on the
dynamics. In particular, it generates an additional
decoherence of the Ramsey signal already at $t=0$, as well as an
enhancement of the oscillations in $|S(t)|$ for resonant
interactions -- see Fig.~\ref{figtwostep}E.

For the calculation of the FDA results shown in Fig.~2 of the
main text we use the same procedure as described above. 
We account for slight additional experimental decoherence
by scaling the prediction for $|S(t)|$ as described in Section S.5C. We also note
that the phase $\varphi_\text{FDA}(t)$ of the Ramsey signal
$S(t)=|S(t)|e^{-i\varphi_\text{FDA}(t)}$, as determined from
Eq.~\eqref{modS}, differs from the experimentally measured phase
$\varphi(t)$ due to the detuning of the rf frequency from
$\omega_0$. They are related by 
$\varphi(t) = \varphi_\text{FDA}(t) - (\omega_\text{rf}-\omega_0) (2
t_\text{wait} + t_\text{rf})$.
Similar to the previous section and to the experiment, we take 
$\omega_\text{rf}-\omega_0 = \varepsilon_1/\hbar$.

As outlined in Section \ref{sec:S3A}, in the idealized Ramsey scenario
the Fourier transform $A(\omega)$ of $S(t)$ is equivalent to the rf
absorption in linear response, cf.~\eqref{ASRelation}
~\cite{Knap2012tdi}. Similarly to our $T=0$ analysis in
Sec.~\ref{sec:rfpulse}, we now study the effect of the residual
interactions $X_1$ on the spectral decomposition of $S(t)$. To this
end we compare the two signals $A(\omega)$ for the perfect quench with
the result obtained for the experimental sequence as modelled
by~Eq.~\eqref{modS}.  We show the comparison of the spectra obtained
in the idealized (dashed) and experimentally realized scenario (solid)
in the lower panel of Fig.~\ref{figtwostep}. As for our $T=0$ results
discussed above, we find only a small difference between the two
finite-temperature spectra. Therefore, in agreement with the
experimental observation, cf. Fig.~3 in the main paper, under the
condition of $|X_1|\approx 5$ we see that the weak interactions during
the rf pulses have an observable but small effect on the predicted
spectra.

In accordance with the results from the TBM shown in
Fig.~\ref{fig:imperfect}, we find from the evaluation of
Eq.~\eqref{modS} that weak interactions $X_1$ lead to a small shift of
spectral weight into the corresponding dominant polaron 
branches. This shift of spectral weight is also observed
experimentally, see Fig.~3 of the main text.

\subsection{Stronger interactions during rf pulses: 
  illustration of quantum state preparation}

\begin{figure*}[thb]
\centering
\includegraphics[width=12cm]{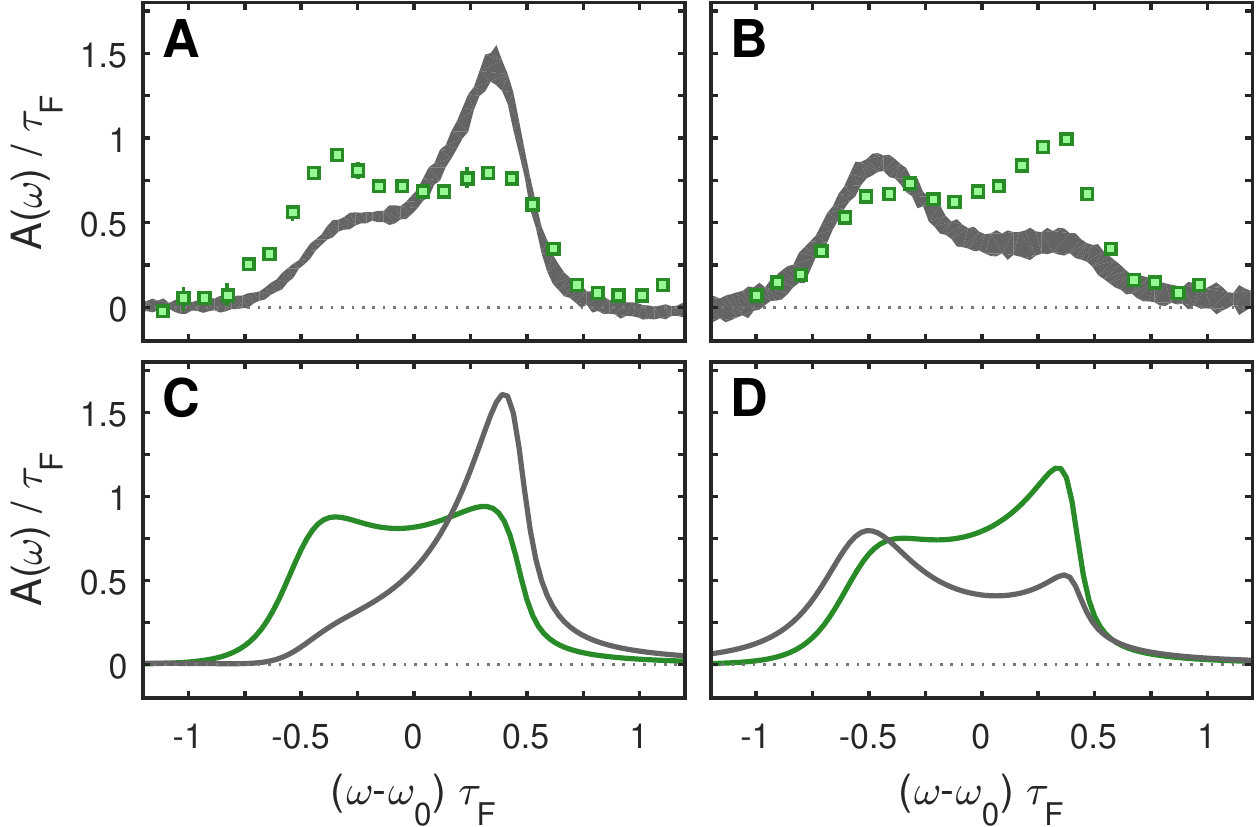}
 \caption{ \textbf{Control of the spectral decomposition of many-body
     quantum states.} Upper panel: We compare the experimentally
   measured rf spectrum at the interaction parameter $X$ (green squares) to
   the Fourier transform of $S(t)$ obtained using the measurement
   procedure illustrated in Fig.~\ref{fig:rfsequence} with initial
   interaction parameter $X_1$ (gray shading). Lower panel: we compare
   the theoretical prediction from the FDA for the linear-response excitation 
   spectrum (green) to the Fourier transform of the signal obtained
   by simulating the experimental sequence according to Eq.~\eqref{modS}.
   ({\bf A}, {\bf C}) $X=0.14$, $X_1=-2.2$, $k_F R^*=1.09$,
   $T/T_F=0.174$. ({\bf B}, {\bf D}) $X=-0.25$, $X_1=1.7$,
   $k_F R^*=1.1$, $T/T_F=0.17$.
\label{figsmallquench}}
\end{figure*}

The shift of spectral weight towards the attractive or repulsive
branches of the spectrum, cf.~Figs.~\ref{fig:imperfect}
and~\ref{figtwostep}, may be interpreted as follows: The residual
interactions present during the initial $\pi/2$ impurity spin rotation
serve to produce an interacting many-body quantum state.  As such,
this procedure can be viewed as an adiabatic preparation of an
attractive or repulsive polaron. 
Compared to the noninteracting state, this polaron has an increased 
wavefunction overlap with the corresponding branch of the strongly interacting system. 
When the system is then quenched into the regime of strong interactions,
the increased overlap results in the corresponding shift of the spectral weight.
An intriguing question is then whether such an approach can
provide a novel way to experimentally control the spectral
decomposition of quantum states.

To investigate this possibility, we increase the interaction during
the $\pi/2$ rotations, corresponding to decreasing $|X_1|$, and determine the effect on $A(\omega)$. In the
upper panel of Fig.~\ref{figsmallquench} we show the 
spectra obtained by linear-response rf spectroscopy (green
squares). Similar to Fig.~3 of the main paper, we compare this result
to the Fourier transform of the Ramsey signal $S(t)$ (gray shading),
as obtained from the experimental sequence described in
Fig.~\ref{fig:rfsequence}. We also compare our experimental result to
the prediction from the FDA, where the dynamics has been modelled as
described by Eq.~\eqref{modS}. As in the main text, we find excellent
agreement between experiment and theory. Indeed, both feature a strong
shift of spectral weight to regions of the spectrum that are
adiabatically connected to the dominant polaron branches at
interaction $X_1$.  Furthermore, when comparing $A(\omega)$ in
Fig.~\ref{figsmallquench}, with the spectrum for $|X_1|\approx5$ in
Figs.~\ref{fig:imperfect} and~\ref{figtwostep}, it is clear that the
amount by which the spectrum is shifted can be controlled by the
strength of the interaction during the rf pulses.  This strongly
supports the assertion that the initial interactions can be used to
precisely control the many-body dynamics.  Our experimental techniques
thus allow for a precise, dynamic control of the spectral
decomposition of quantum states in future experiments.

The excellent agreement between theory and experiment also
demonstrates that our theoretical approaches can be used to explore
experimental ramps in combination with interferometric protocols in order to
find, for instance, optimized spin and interaction trajectories.

\section{Universal features of impurity dynamics and relation to
 orthogonality catastrophe}

For impurities localized in space, which, for instance, 
can be achieved by species-selective three dimensional optical lattices, 
our experimental setup allows one to study universal features exhibited by the Anderson orthogonality catastrophe \cite{Anderson1967ici}.
The orthogonality catastrophe was originally
studied in the context of x-ray absorption spectra in metals, where
high-energy x-ray photons create atomic core holes by photoemission of
inner-shell electrons \cite{Mahan1990mpp}. These core holes produce a
scattering potential for the electrons in the conduction band, leading
to characteristic power-law edges in the absorption spectra with an
exponent that is universally determined by the scattering phase shift
at the Fermi surface \cite{Anderson1967ici}. However, impurities,
phonons, residual interactions between the electrons, and a lack of
knowledge of microscopic parameters makes it difficult to
unambiguously determine the universal features of the orthogonality
catastrophe in typical solid state materials \cite{Ohtaka1990tot}. In
contrast, the Hamiltonian in our experiment is well characterized on
all relevant energy scales, and therefore the full dynamic response of
the system can be reliably calculated by theory and probed by the
ultrafast experimental techniques demonstrated in this work.  This
enables one to obtain fundamental insights into universal features of
the orthogonality catastrophe, which are difficult to access in other
systems.

\begin{figure*}[tbh]
        \centering
        \includegraphics[width=16cm]{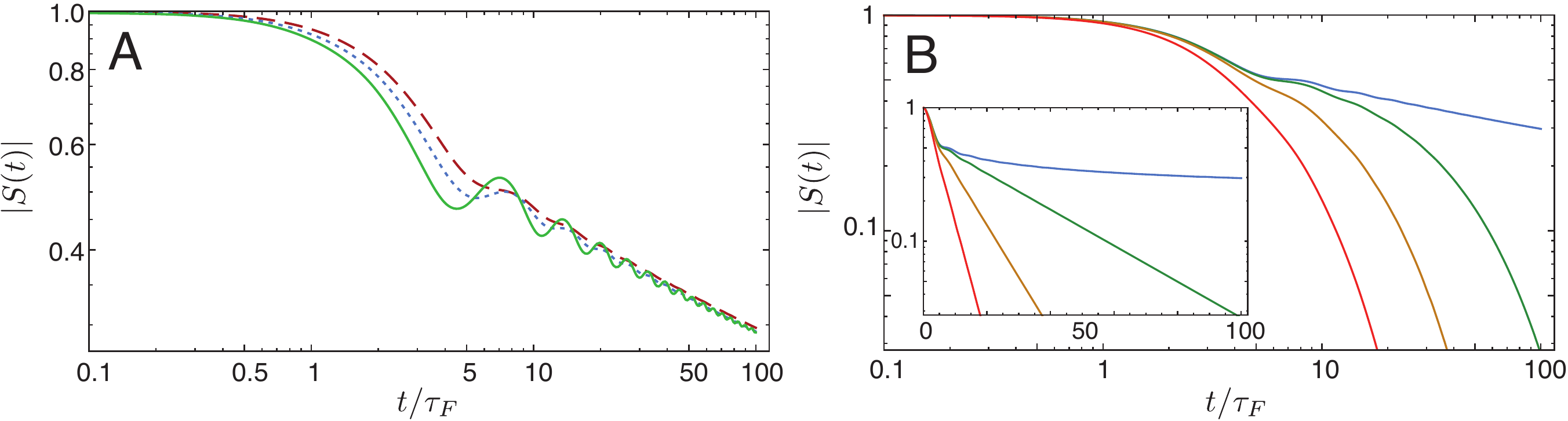}
        \caption{\textbf{Universal features of the dynamical
            orthogonality catastrophe}.  We show the Ramsey contrast
          for an infinitely heavy impurity obtained within the
          FDA. ({\bf{A}}) The zero-temperature Ramsey contrast exhibits a
          power law decay, shown on a double logarithmic scale. We
          change the Feshbach resonance range $k_F R^*$ and
          interaction parameter $X$ in such a way that the scattering
          phase shift at the Fermi surface is constant leading to a
          constant exponent of the power law tail. The data
          corresponds to a fixed phase shift $\delta_{k_F}=1.4$ with
          the choices $(X,k_F R^*)=(1,1.12)$ (dashed red),
          $(X,k_F R^*)=(0.58,0.56)$ (dotted blue), and
          $(X,k_F R^*)=(0.15,0)$ (solid green).  ({\bf{B}}) Ramsey contrast
          at various temperatures on a double logarithmic scale. We
          choose temperatures $T/T_F=0$ (blue), $0.05$ (green), $0.15$
          (orange), $0.4$ (red) at fixed values $X=1$ and
          $k_FR^*=1.12$. The inset shows the same data on a
          logarithmic-linear scale to emphasize the appearance of
          exponential tails at finite temperature.}
        \label{fig:OC} 
\end{figure*}

To illustrate how the orthogonality catastrophe would manifest itself
in an ultracold atomic gas experiment, the response of infinite mass
impurities calculated using the FDA for the perfect quench scenario is
shown in Fig.~\ref{fig:OC}.  First, at short times and for a range parameter of the Feshbach resonance $R^*>0$, we
see that the Ramsey contrast decays quadratically for all scattering
parameters and temperatures considered, in accordance with
\Eq{eq:shortt}. The main universal feature associated with the
orthogonality catastrophe is expected in the long-time dynamics at
$T=0$: Here, the Ramsey response is predicted to exhibit power law
tails, which depend only on the scattering phase shift at the Fermi
surface \cite{Anderson1967ici,Knap2012tdi}. This is explicitly
verified in Fig.~\ref{fig:OC}A where we fix the scattering phase
shift at the Fermi surface but change the scattering parameters. While
the response at intermediate times depends on the scattering
parameters, we see that the long-time evolution approaches a
universal power law that only depends on the phase shift at the
Fermi surface. We note that the long-time dynamics is universal:
It is the same for a system with a broad resonance where $R^*=0$
(solid line in Fig.~\ref{fig:OC}A), as it is for our system with a
finite range parameter (dashed and dotted lines).

When the temperature is non-zero, as in the experiment, thermal
fluctuations alter the power law dephasing dynamics at sufficiently
long times. Instead, exponential tails due to thermal decoherence
appear as another universal feature of the
dynamics~\cite{Korringa1950nmr,Anderson1967ici,Yuval1970erf,Knap2012tdi}. The
exponential tails are illustrated in Fig.~\ref{fig:OC}B.
The effects of thermal decoherence could be countered by
employing the recently developed cooling methods \cite{Hart2015ooa},
opening the door to observing the orthogonality catastrophe in a cold-atom system.

Finally, we note that in our experiment temperature becomes relevant
at a time scale similar to those associated with recoil and multiple
particle-hole excitations.  It is a challenge for theoretical
approaches to exactly account for both recoil and higher order
particle-hole excitations \cite{Rosch1999qct}. However, experiments at
lower temperatures which take advantage of the tunability of the
impurity mass using optical lattices would be ideally suited to probe
the competition between these effects. Such ultracold-atom 
experiments would hence provide important insight into this long
standing theoretical question.

\section{Experimental and data analysis procedures}

In this section we discuss the procedures used to record and analyze the data presented in this work.
We detail the cooling and preparation of our atomic samples, the details of the rf pulses used in our Ramsey sequences,
the methods used to analyze the data and the method that we use to vary the concentration of the K atoms.

\subsection{Sample preparation}

The atomic samples are prepared by forced evaporation of Li atoms from a Li-K mixture held in an optical trap, where the K atoms are sympathetically cooled by the Li environment. This preparation procedure is described in detail in Refs.~\cite{Trenkwalder2011heo,Spiegelhalder2010aop}.
At the end of the forced evaporation, the Li and K atoms are transferred into an optical trap composed of two crossed 1064-nm laser beams, as described in Ref.~\cite{Cetina2015doi}. 
The measured radial and axial trap frequencies of the Li atoms are $f_{r,\rm{Li}}=941(5)$~Hz and $f_{z,\rm{Li}}=134(1)$~Hz, respectively.
The measured radial and axial trap frequencies of the K  atoms are $f_{r,\rm{K}}=585(3)$~Hz and $f_{z,\rm{K}}=81(1)$~Hz, respectively.

At the end of the preparation procedure, the Li and the K atoms are in their lowest Zeeman states Li$|1\rangle$ and K$|1\rangle$.
Before the Ramsey sequence, the K atoms are transferred to the K$|2\rangle$ state using an rf pulse. 
Following this rf transfer, the Li and K atoms are thermalized by holding them for 750 ms in the crossed-beam optical trap.
While the interaction between the Li$|1\rangle$ and K$|2\rangle$ atoms, characterized by the scattering length $a_{12}=63 a_0$ \cite{Naik2011fri}, 
is sufficient to ensure thermalization during this hold time, it can be neglected during the Ramsey experiments.
The temperature of the atoms is determined by releasing the atoms from the trap and observing the free expansion of the K cloud.

Due to the Li Fermi pressure and the more than two times stronger optical
potential for K, the K cloud is much smaller than the Li cloud \cite{Trenkwalder2011heo},
and therefore samples a nearly homogeneous Li environment. 
Because of the small variation of the Li environment sampled by the K atoms, 
we introduce the effective Li Fermi energy $\epsilon_F$ as
\begin{align}
\epsilon_F =\frac{1}{N_{\rm{K}}}\int E_F({\bf{r}}) n_{\rm{K}}({\bf{r}}) d^3 {\bf{r}}\, .
\end{align}
Here, $n_{\rm{K}}({\bf{r}})$ is the local K number density at position $\bf{r}$ in the trap,
and 
\begin{align}
E_F({\bf{r}})=\frac{\hbar^2\left(6\pi^2 n_{\rm{Li}}({\bf{r}})\right)^{2/3}}{2 m_{\rm{Li}}}
\end{align}
is the local Li Fermi energy as determined by the local Li number density $n_{\rm{Li}}({\bf{r}})$.
We quantify the small inhomogeneity of the Li environment experienced by the K atoms by the standard deviation of the local Li Fermi energy 
\begin{align}
\sigma(E_F)=\left(\frac{1}{N_{\rm{K}}}\int (E_F({\bf{r}})-\epsilon_F)^2 n_{\rm{K}}({\bf{r}}) d^3 {\bf{r}}\right)^{1/2}\, .
\end{align}
We also introduce the average Li and K number densities $\bar{n}_{\rm{Li}}$ and $\bar{n}_{\rm{K}}$ sampled by the K atoms as
\begin{align}
\bar{n}_{\rm{Li,K}} = \frac{1}{N_{\rm{K}}}\int n_{\rm{Li,K}}({\bf{r}}) n_{\rm{K}}({\bf{r}}) d^3 {\bf{r}}\,.
\end{align}

In contrast to the Li atoms, the K atoms in our measurements remain non-degenerate, with $k_B T/E_F^{\rm{K}}(0) > 1.2$, where
$E_F^{\rm{K}}(0)$ is the local potassium Fermi energy in the center of the trap when all K atoms are in the same internal state.

\begin{table*}
\begin{tabular}{|c|c|c|c|c|c|c|c|c|}
  \hline
  Figure(s) & $N_\mathrm{Li}$	& $N_\mathrm{K}$	& $T$		& $\epsilon_F/h$	& $\frac{\sigma(E_{F})}{\epsilon_F}$ & $\bar{n}_\mathrm{Li}$	& $\bar{n}_\mathrm{K}$	\\ 
			& $(10^5)$ 				& $(10^4)$ 				& (nK)	& (kHz) 									& \% & $10^{12}$cm$^{-3}$		& $10^{12}$cm$^{-3} $ 	\\
  \hline
  \hline
	2A, 2C, 3A & 3.5(4)	& 0.95(10)  & 435(25)	& $54.6(2.7)$ & 7.4 & 8.9(7)	& 1.8(3) \\ 
  \hline
	2B, 2D, 3B & 3.3(4)	& 1.0(1)		& 410(25)	& $53.9(2.4)$	& 7.1   & 8.7(6) 	& 2.0(3)  \\ 
  \hline
	2E, 2F, 3C  & 3.5(4)  & 1.0(1) 	& 460(30)	& $54.1(2.4)$	& 7.7   & 8.8(6)	& 1.7(3) \\ 
  \hline
	\ref{figsmallquench}A & 3.1(4) & 1.0(1)	  & 430(30)	& $52.0(2.9)$ & 	7.7	& 8.2(7) 	& 1.8(3) \\ 
  \hline
	\ref{figsmallquench}B & 2.9(3) & 1.05(10)	& 425(35)	& $50.8(2.1)$ & 	7.7	& 8.0(6) 	& 2.0(3)  \\
  \hline
	4  & 2.35(30)	& 2.5(1)	& 520(25)	& $44.2(2.3)$ 		& 10.4  & 6.5(5) 	& 3.4(3)  \\
\hline
\end{tabular}
\caption{The total number of the Li atoms $N_{\rm{Li}}$, the total number of the K atoms $N_{\rm{K}}$, the sample temperature $T$, the effective Li Fermi energy $\epsilon_{F}$, the standard deviation $\sigma(E_F)$ of the local Li Fermi energy across the trap, the trap-averaged Li and K number densities $\bar{n}_{\rm{Li}}$ and $\bar{n}_{\rm{K}}$ 
in our measurements.}
\label{table:ExpSettings}
\end{table*}

For all measurement presented in this work, Table \ref{table:ExpSettings} lists the total numbers of the Li and K atoms, their temperatures and trap-averaged densities, as well as the effective Li Fermi energies and their standard deviations. 
Throughout our measurements, these parameters remain nearly constant, with the exception of the measurements shown in Fig.~4.
Here, in order to investigate the effect of the K concentration, the total number of the K atoms is increased from about $1\times10^4$ to $2.5\times10^4$.
The attendant increase in the thermal load during the Li evaporation results in a decrease of the Li atom number and
an increase in the temperature of the final atomic sample.

Note that, in contrast to our previous work \cite{Cetina2015doi}, our present experiments have been optimized for large optically induced interaction shifts ($|X-X_1|\approx5$). 
These shifts are produced by switching one of the crossed trapping beams from a beam with a low peak intensity and small size to a beam with a large intensity and large size  propagating in the same direction. In our previous work \cite{Cetina2015doi}, as well as in the measurements shown in Fig.~\ref{figsmallquench}, 
the waists, positions and intensities of the two beams are adjusted so as to yield mode-matched trapping potentials, preventing excitations of the center-of-mass and breathing collective modes of the atomic clouds. In the measurements presented in Figs. 2, 3 and 4, a larger beam intensity was used in order to produce a larger optical shift, resulting in some 
excitation of the breathing modes.

The maximal interaction time in our Ramsey measurements of 60 $\mu$s is much smaller than the shortest period of a collective oscillation (about 500 $\mu$s).
We calculate that, during our short interaction time, the oscillations of the breathing modes cause at most a 6\% variation of $\epsilon_F$ around its initial value specified in Table \ref{table:ExpSettings}, 
without any significant effect on the measurements presented here.

\subsection{Rf pulses}


We apply rf pulses in the Ramsey procedures by discretely gating a continously running rf source. 
To record the atomic populations $N_3$ and $N_2$ as a function of the phase of the second rf pulse, we change the phase of the rf source by a variable amount $\phi_{\rm{rf}}$
before applying this pulse.

The weak interactions between the K atoms in the K$|3\rangle$ state and the Li atoms corresponding to the interaction parameter $X_1$ cause the transition frequency between the K$|2\rangle$ and the K$|3\rangle$ states
to differ from the transition frequency $\omega_0$ in the absence of the Li atoms. 
We compensate for this effect by adjusting the frequency $\omega_{\rm{rf}}$ of the rf source
to be resonant with the K$|2\rangle$$-$K$|3\rangle$ transition at the time when the rf pulses are applied.
For the data in Figs.~2A, 2B, 2C, $(\omega_{\rm{rf}}-\omega_0)\tau_F$ is equal
to $+0.06$, $-0.07$, $-0.05$, respectively. 
For the data in Fig.~\ref{figsmallquench}C and \ref{figsmallquench}D where the interaction of the K atoms during the rf pulses is stronger,
$(\omega_{\rm{rf}}-\omega_0)\tau_F$ is equal to $+0.11$ and $-0.16$.

The shift in the frequency of the rf source from $\omega_0$ to $\omega_{\rm{rf}}$ causes the signal $S(t)$ to accumulate an additional phase 
$(\omega_{\rm{rf}}-\omega_0)t$ during the interaction time $t$. To account for this added phase, we introduce the phase
$\phi=\phi_{\rm{rf}}+(\omega_{\rm{rf}}-\omega_0)t$.

\subsection{Analysis methods}

We determine the contrast $|S(t)|$ and the phase $\varphi(t)$ by fitting
the Ramsey signal $(N_3-N_2)/(N_3+N_2)$ as a function of the phase $\phi$ 
to a sine wave with an offset i.e. $F(t)+|S(t)|\cos\left(\phi-\varphi(t)\right)$.
Decoherence during the rf pulses, as well as imperfections of the rf pulses
and the atom detection, cause the contrast for $t=0$ to be slightly smaller than unity. When comparing
theoretical results from Figs.~\ref{fig:imperfect} and \ref{figtwostep}  
to the experimental data in Fig. 2, we account for this effect by
scaling the theoretical predictions for $|S(t)|$ by an overall factor $\eta$.
For each calculation, this factor is determined by fitting the prediction for $|S(t)|$ to the three data points 
with the the shortest interaction times. 
We obtain $0.92<\eta<1$, which corresponds to an additional loss of contrast that is of the same order as the 
decoherence during the rf pulses predicted by the FDA (see Fig. \ref{figtwostep}).

To compute the Fourier transform of the experimental $S(t)$ data, we use piecewise linear interpolations 
of $\log S(t)$ and $\varphi(t)$ between the individual data points. Outside of the range of the data, we set $S(t)=0$. 
To determine the error of the Fourier transform, we sample
the values of $S(t)$ and $\varphi(t)$ at each data point 
from Gaussian distributions whose means and standard deviations 
correspond to the measured values and errors, respectively.
We use the standard deviation of the computed values of the Fourier transform 
for each value of $\omega$ as an estimate of the error indicated by the shaded
areas in Figs. 3 and \ref{figsmallquench}.

\subsection{Varying the K concentration}

We study the effects of the impurity concentration by varying the number of the strongly interacting K atoms.
If this were done by changing the total number of the K atoms in the experiment, the change in the thermal load on the Li atoms during forced evaporation
would result in a correlated variation in the number of Li atoms and the sample temperature (compare the settings for Fig.~2 and Fig.~4 in Table \ref{table:ExpSettings}).
To avoid these systematic effects, in the measurements presented in Fig.~4, we keep the total number of the K atoms constant
and vary the fraction of the K atoms that participate in the Ramsey sequence.
We accomplish this by changing the intensity of the rf pulse that transfers the K atoms from the $|1\rangle$ state to the
$|2\rangle$ state before the Ramsey procedure. During the subsequent 750 ms preceding the
Ramsey sequence, the K atoms collisionally thermalize with the much larger Li cloud, resulting
in an incoherent mixture of K$|1\rangle$ and K$|2\rangle$ atoms at a constant temperature.
When referring to these measurements, we use $\bar{n}_{\rm{K}}$ not for the average density of all K atoms, but for the density of those K atoms that participate in the Ramsey sequence.

We minimized the small effects of long-time drifts in the temperature, the atom numbers and the trapping potential
by varying the experimental parameters in a specific order.
For each K concentration and interaction time, we recorded data for 4 different phases of the second rf pulse in order to obtain $S(t)$.
For each interaction time, the data with different K concentrations were recorded in immediate succession. 
The data sets for different interaction times were then recorded in a random order.


\section{Linearity of rf response}

The response of atoms to an applied rf field is linear if the fraction of the atoms transferred from one state to another is proportional to the intensity of the field. 
Linearity can be ensured by using a sufficiently weak rf pulse that is also much longer than the inverse width of the relevant spectral features.
The narrowest spectral features in the present work are the polaron peaks in Figs.~3A and 3B with rms widths $0.06\,\hbar/\tau_F$ and $0.09\,\hbar/\tau_F$, respectively.
To record these polaron spectra, we used Blackman-shaped rf pulses \cite{Kasevich1992lcb} whose duration $t_{\rm{rf}}=300\,\mu{\rm{s}}\approx100\,\tau_F$
is much longer than the inverse widths of the polaron peaks.

\begin{figure*}[th]
        \centering
        \includegraphics[width=14cm]{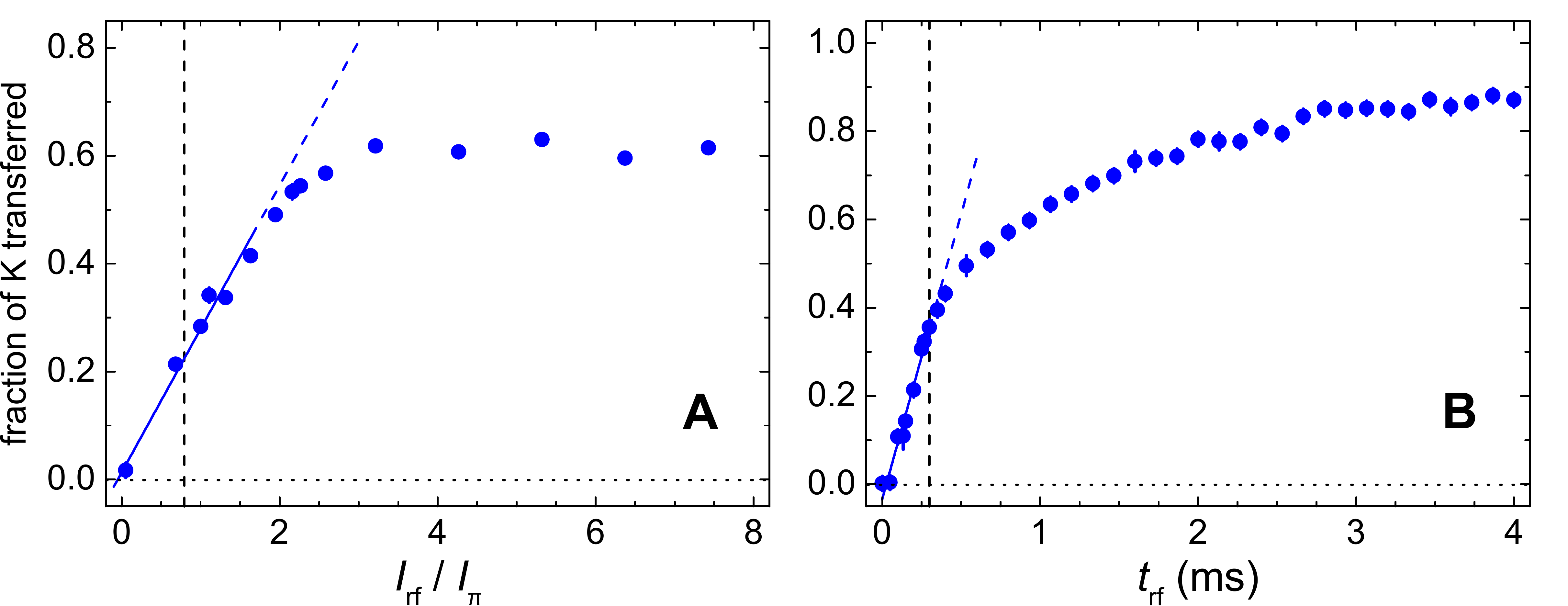}
        \caption{\textbf{Linearity of the rf response in the repulsive polaron regime}. (\textbf{A}) Fraction of the K atoms transferred from state K$|2\rangle$ to the state K$|3\rangle$ for $X=-0.13(6)$ as a function of the intensity $I_{\rm{rf}}$ of an rf pulse with duration $t_{\rm{rf}}=300\,\mu$s. (\textbf{B}) Fraction of the K atoms transferred for $X=-0.23(6)$ as a function of the duration $t_{\rm{rf}}$ of the rf pulse for the rf pulse intensity $I_{\rm{rf}}=0.79\, I_{\pi}$. Vertical dashed lines correspond to $I_{\rm{rf}}=0.79\,I_{\pi}$ and $t_{\rm{rf}}=300\,\mu$s, respectively. The pulse frequencies are adjusted to resonantly excite the repulsive polaron. The blue solid lines indicate linear fits to the data in the ranges indicated by the same lines. The blue dashed lines show extrapolations of these fits. 
        }
        \label{fig:rflin} 
\end{figure*} 

\begin{figure}[th]
        \centering
        \includegraphics[width=8cm]{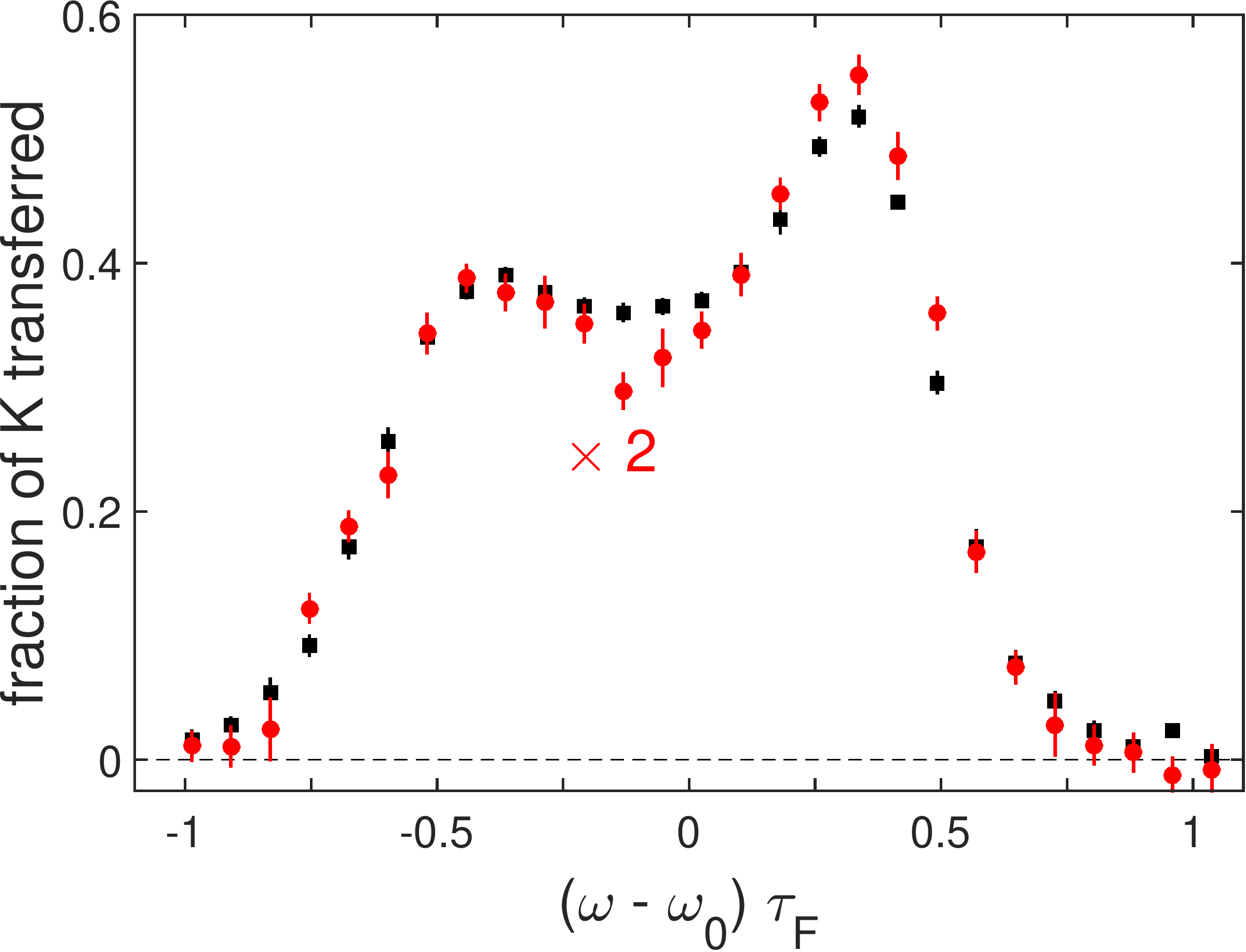}
        \caption{\textbf{Linearity of the rf response for resonant interactions}. Fraction of the K atoms transferred from state K$|2\rangle$ to the state K$|3\rangle$ by an rf pulse with duration $t_{\rm{rf}}=100\,\mu$s for $X=+0.02(6)$. For the black data points, the intensity of the rf pulse is adjusted to obtain a $\pi$-pulse in the absence of Li atoms. The red data points correspond to a 50\% lower intensity of the rf field.
        }
        \label{fig:reslin} 
\end{figure} 

We checked the linearity of the response by varying the intensity $I_{\rm{rf}}$ of the applied rf field. 
Fig.~\ref{fig:rflin}A shows the fraction of the K atoms transferred from the K$|2\rangle$ to the K$|3\rangle$ state 
in the repulsive polaron regime, under conditions similar to those in the measurements shown in Fig.~3A.
The frequency of the rf pulse is adjusted so that $(\omega_{\rm{rf}}-\omega_0)\tau_F=0.3$, corresponding to peak response and resonant excitation of the repulsive polaron.
The rf intensity is measured in units of the intensity $I_{\pi}$ that results in a $\pi$-pulse for noninteracting K atoms.
For intensities up to the intensity $I_{\rm{rf}}=0.79~I_{\pi}$, which is used in the measurements shown in Figs.~3A and 3B,
we observe that the transferred fraction of the K atoms stays essentially proportional to the intensity of the pulse.

In the linear-response regime, the atomic response is predicted to be proportional to the duration of the rf pulse.
Fig.~\ref{fig:rflin}B shows the fraction of the K atoms transferred in the repulsive polaron regime by rf pulses with $I_{\rm{rf}}=0.79~I_{\pi}$, 
as a function of the pulse duration. The frequency of the rf pulse is adjusted so that $(\omega_{\rm{rf}}-\omega_0)\tau_F=0.3$,
in order to obtain the peak response, as in Fig~\ref{fig:rflin}A. 
For pulses with duration up to 300 $\mu$s (indicated by the dashed line),
we observe that the transferred fraction of the K atoms stays essentially proportional to the duration of the pulse.

Note that the maximal transferred fraction exceeds 0.5.
We explain this observation by the coupling of the initial non-interacting K state to multiple interacting K states by the rf pulse,
which manifest themselves as the polaron peak and the broad pedestal in our spectra.

The spectra for resonant Li-K interactions shown in  Figs.~3B, \ref{figsmallquench}A, \ref{figsmallquench}B were recorded using Blackman-shaped rf pulses 
with duration of $t_{\rm{rf}}=100\,\mu$s (approximately $35\,\tau_F$). 
The intensity of these pulses was adjusted to 50\% of that needed to produce $\pi$ pulses for noninteracting K atoms. We verified the linearity of the rf response by comparing the spectra recorded using this rf intensity to those recorded using the intensity needed to produce full $\pi$ pulses for noninteracting K atoms (Fig.~\ref{fig:reslin}).
Our observations are in good agreement with linear response.

\bibliographystyle{apsrev}
\bibliography{ultracold,library,fastdyn}

\end{document}